\documentclass[10pt,twocolumn,letterpaper]{article}

\usepackage[final]{cvpr}      % To produce the REVIEW version
\usepackage{algorithm}

\usepackage{subcaption}

\usepackage[noend]{algpseudocode}  
\usepackage{amsmath}
\usepackage{booktabs}
\DeclareMathOperator*{\argmin}{arg\,min}

\usepackage{xr}
\usepackage{multirow}
\usepackage{tabularx}
\newcolumntype{C}{>{\centering\arraybackslash}X}

\definecolor{cvprblue}{rgb}{0.21,0.49,0.74}
\usepackage[pagebackref,breaklinks,colorlinks,allcolors=cvprblue]{hyperref}

\title{Robust Client-Server Watermarking for Split Federated Learning}

\author{
    Jiaxiong Tang\thanks{Equal contribution.} \quad
    Zhengchunmin Dai\footnotemark[1] \quad
    Liantao Wu\thanks{Corresponding author.} \quad
    Zhenfu Cao\\
    East China Normal University \\
    {\tt\small 51275902027@stu.ecnu.edu.cn \quad 
    51275902069@stu.ecnu.edu.cn \quad 
    ltwu@sei.ecnu.edu.cn \quad
    zfcao@sei.ecnu.edu.cn}
    \and
    Peng Sun\\
    Hunan University\\
    {\tt\small psun@hnu.edu.cn}
    \and
    Honglong Chen\\
    China University of Petroleum\\
    {\tt\small chenhl@upc.edu.cn}
}

\begin{document}
\maketitle
\begin{abstract}
Split Federated Learning (SFL) is renowned for its privacy-preserving nature and low computational overhead among decentralized machine learning paradigms. In this framework, clients employ lightweight models to process private data locally and transmit intermediate outputs to a powerful server for further computation. However, SFL is a double-edged sword: while it enables edge computing and enhances privacy, it also introduces intellectual property ambiguity as both clients and the server jointly contribute to training. Existing watermarking techniques fail to protect both sides since no single participant possesses the complete model.
To address this, we propose RISE, a \textbf{R}{obust} model \textbf{I}{ntellectual} property protection scheme using client-\textbf{S}{erver} watermark \textbf{E}{mbedding} for SFL. Specifically, RISE adopts an asymmetric client–server watermarking design:
the server embeds feature-based watermarks through a loss regularization term, while clients embed backdoor-based watermarks by injecting predefined trigger samples into private datasets.
This co-embedding strategy enables both clients and the server to verify model ownership. Experimental results on standard datasets and multiple network architectures show that RISE achieves over \(95\%\) watermark detection rate (\(p-value<0.03\)) across most settings. It exhibits no mutual interference between client- and server-side watermarks and remains robust against common removal attacks. 
Our source code is available at \href{https://github.com/dickton/RISE}{\includegraphics[height=2.0ex]{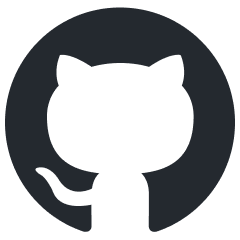}}.
% Our source code is available at (\url{https://anonymous.4open.science/r/RISE-66D3/}).

\end{abstract}    
\section{Introduction}

Federated Learning (FL)~\cite{9084352} has become a cornerstone paradigm for privacy-preserving distributed machine learning, enabling multiple clients to collaboratively train a global model without exposing their local data. 
By decentralizing model updates rather than data collection, FL mitigates direct privacy leakage and has been widely adopted in domains such as healthcare~\cite{9879840, Chen_2024_CVPR, Xie2025dFLMoE}, finance~\cite{fedFinance} and recommendation systems~\cite{WAHAB2022189, NEURIPS2024_774164b9, Wang2025FedReccommend}.

% However, the distributed nature of FL significantly expands the attack surface, as both clients and servers jointly participate in training across potentially untrusted environments. This opens the door to adversarial behaviors such as model extraction or inversion, leading to unauthorized use or monetization of proprietary models~\cite{10468659}. 
% To mitigate these risks, model watermarking has emerged as a promising defense mechanism~\cite{263804, 9645219, 9835648}. Inspired by multimedia digital watermarking, these techniques embed distinct, identifiable watermarks, such as bit strings or backdoors, into models during training. Once embedded, these watermarks can be later detected to verify ownership without compromising model performance.

However, the decentralized nature of FL also broadens the attack surface, as both clients and servers jointly participate in training across potentially untrusted environments. This exposes models to risks such as extraction and inversion attacks, leading to unauthorized redistribution or monetization of proprietary models~\cite{10468659, Wang2022freerider}. 
To counter these threats, model watermarking has emerged as an effective approach for verifying model ownership~\cite{263804,9645219,9835648}. These methods embed distinct, identifiable patterns—such as bit strings or backdoors—into model parameters during training, which can later be detected to confirm legitimate ownership without affecting model performance.

% However, the distributed nature of FL poses significant intellectual property risks \cite{10888286,10376357}. As both clients and servers participate in training, adversaries may exploit vulnerabilities to extract or steal proprietary models, leading to unauthorized usage or monetization \cite{10468659}. To mitigate these risks, model watermarking has emerged as a promising defense mechanism \cite{263804,9645219,9835648}. Inspired by multimedia digital watermarking, these techniques embed distinct, identifiable watermarks, such as bit strings or backdoors, into models during training. Once embedded, these watermarks can be later detected to verify ownership without compromising model performance.

%Despite these advantages, the distributed nature of FL introduces new security and intellectual property risks \cite{10888286,10376357}. Since clients and servers collaboratively train the model, malicious parties may attempt to extract or steal proprietary models, leading to unauthorized use or monetization \cite{10468659}. To address these urgent concerns, model watermarking techniques have emerged as a promising solution \cite{263804,9645219,9835648}. Drawing inspiration from multimedia digital watermarking, model watermarking techniques embed special identifiable watermarks such as bit string or backdoor during training. These embedded watermarks can subsequently be detected to establish legal ownership claims without compromising model functionality.

Although model watermarking methods have been extensively studied in centralized learning~\cite{Uchida2017watermarkDNN, DNNwatermark1, DNNwatermark2}, researchers have developed specialized watermarking algorithms for FL due to its decentralized nature and the inability of the central server to access training data~\cite{10437312}. The FL-specific watermarking techniques can be broadly classified into server-side~\cite{CHEN2023103504} and client-side approaches~\cite{RN205}. Server-side watermarking embeds watermarks into the global model after aggregation at the server, which is then distributed to clients for local training. In contrast, client-side watermarking involves embedding watermarks into local model parameters during training, with the watermarked parameters subsequently submitted for aggregation.

Yet, existing server-side or client-side FL watermarking techniques cannot be directly applicable to Split Federated Learning (SFL)~\cite{thapa2022splitfed, SFL2, 10740645}. SFL integrates the collaborative training framework of FL with the model-splitting strategy of split learning (SL)~\cite{gupta2018distributed}, 
enabling efficient training while preserving data locality and alleviating the computational burden on resource-constrained edge devices. By partitioning the neural network vertically into client-side and server-side models\footnote{Throughout this paper, we refer to the client-side model as the bottom model, and the server-side model as the top model.},  SFL reduces the computational burden on edge devices and enables the decentralized training of large models. As both clients and the server actively contribute to model training, securing the model’s intellectual property requires a more comprehensive watermarking strategy, which faces the following two challenges: 

\textbf{Absence of SFL-specific watermarking scheme.}  Due to the unique architecture of SFL, neither the client nor the server has full access to the entire model or complete control over the training process. Consequently, existing client-side and server-side FL watermarking schemes are not directly applicable to SFL. A joint client-server watermarking scheme~\cite{RN212} for FL is proposed, where a backdoor-based watermark is embedded on the server side, and a feature-based watermark is embedded on the client side. However, in SFL, clients lack direct access to most of model parameters managed by the server, while the server cannot control clients’ raw data inputs or labels. As a result, this approach cannot be directly applied to SFL.

% \textbf{Challenge 2: }
\textbf{Designing an Effective Dual-side Watermarking Scheme.} Due to the unique architecture of SFL, both the server and the client contribute to training. Therefore, a watermarking scheme must be designed to allow both parties to simultaneously claim ownership of the model while preserving its effectiveness. Specifically, the watermark should achieve a high detection success rate while maintaining the accuracy of the model's primary task. Furthermore, the watermark 
 should exhibit robustness against removal attacks and prevent interference between the dual-end watermarks. These requirements present significant challenges for the design of an effective watermarking scheme.

Motivated by the aforementioned discussion, we propose RIPS, a \textbf{R}obust model \textbf{I}ntellectual property protection scheme using \textbf{S}erver-client watermark \textbf{E}mbedding for SFL. RIPS enables both clients and the server to verify model ownership. The scheme compRIPSs two key phases: co-embedding and verification. In the co-embedding phase, watermarks are synchronously embedded on both sides during the training process: (1) The server embeds feature-based watermarks into top-layer parameters by incorporating a watermarking regularization term into the main loss function, leveraging its control over critical model parameters and the training process. (2) Clients embed backdoor-based watermarks by injecting predefined trigger samples with poisoned labels into their private datasets, utilizing each client’s autonomy over its data. In the verification phase, both the server and clients can verify their ownership of the trained SFL model. In summary, the main contributions of this paper are as follows:

\begin{itemize}
  % \item To the best of our knowledge, we propose the first dedicated ownership verification scheme for SFL enabling both clients and the server to verify model ownership.
  \item To the best of our knowledge, this work presents the first ownership verification framework specifically designed for SFL, enabling both clients and the server to independently verify model ownership.

  % \item We design a joint client-server watermark embedding mechanism to address the dual-side ownership requirement. This architecture enables independent watermark embedding on both sides: the server embeds feature-based watermarks into server-side model by modifying the loss function, while clients embed backdoor-based watermarks into their local models by injecting predefined triggers into their datasets.
  \item We develop a joint client–server watermarking mechanism to address the dual-side ownership requirement in SFL. This architecture supports independent watermark embedding on both sides: the server embeds feature-based watermarks into the server-side model via a loss regularization term, whereas clients embed backdoor-based watermarks into their local models by injecting predefined triggers into private datasets.

  % \item Experimental results demonstrate that our framework achieves state-of-the-art performance in fidelity, watermark statistic significance, and robustness against removal attacks, including fine-tuning, pruning, and quantization attack.
  \item Experiments demonstrate that RIPS achieves state-of-the-art performance in model fidelity, watermark detectability significance, and robustness against common removal attacks, including fine-tuning, pruning, quantization and Neural Cleanse.
\end{itemize}
\section{Preliminaries and Related Works}

\subsection{Preliminaries}\label{sec:main_task}

\noindent In SFL, each client \(k \in \{1,\dots,K\}\) holds a private dataset \(D_k=\{(x_i,y_i)\}_{i=1}^{n_k}\) and a lightweight bottom model \(B_k\), while the server maintains a larger top model \(T\). During training, client \(k\) forwards smashed data \(z_k = B_k(x_k)\) and labels to the server, receives the gradient \(g_k = \partial L_k / \partial z_k\), and updates \(B_k\) accordingly. The server predicts \(\hat{y}_k=T(z_k)\) and computes the loss \(L_k=\mathcal{L}(T(z_k),y_k)\). 

\noindent The objective of SFL is to jointly learn \(\{B_k\}_{k=1}^{K}\) and \(T\) such that \(T(B_k(x))\) accurately predicts \(y\):
\begin{equation}
\label{eq:main_task_formalization}
    \argmin_{\{B_k\},T}
    \frac{1}{K}\sum_{k=1}^{K}\mathbb{E}_{(x,y)\in D_k}
    [\mathcal{L}(T(B_k(x)),y)].
\end{equation}
 
% While SFL enables privacy-preserving collaborative training, its structural separation causes an inherent ambiguity in model ownership, as neither clients nor the server hold the complete model—motivating our dual-side watermarking design.

This model-splitting mechanism not only preserves data privacy but also offloads most of the computational workload to the more powerful server. The left part of Fig.~\ref{fig:sfl_framework} illustrates a typical SFL workflow.

\subsection{FL Watermarking}
\noindent Since no watermarking schemes have been designed specifically for SFL, we next review the watermarking mechanisms developed for FL, the distributed training paradigm most closely related to SFL. Existing FL watermarking methods aim to protect the intellectual property of the trained models and can be broadly categorized into server-side and client-side techniques.

% \subsection{Split Federated Learning}
% \noindent 

\subsubsection{Server-side Watermarking}

\noindent WAFFLE~\cite{RN205} is the first FL watermarking scheme. It is a server-side scheme comprising two key steps: pretraining and retraining. After each aggregation of local models on the server, backdoor watermarks are embedded into the global model. Additionally, WAFFLE introduces a data-independent trigger set generation method, which has since become the baseline for many subsequent server-side backdoor watermarking approaches.
In contrast, a feature-based watermarking scheme~\cite{RN375} was proposed to sequentially embed client-submitted watermarking strings. While this approach assumes a fully trusted server, it emphasizes the role of client contributions in the model’s development. Furthermore, a hybrid watermarking mechanism~\cite{RN206} was introduced to enhance both ownership verification and traceability. This scheme integrates a global backdoor watermark with client-specific feature-based watermarks and leverages continual learning to preserve model performance on both the primary and watermarking tasks.
% In contrast, \cite{RN375} proposed a feature-based watermarking scheme that sequentially embeds client-submitted watermarking strings. While their approach assumes a fully trusted server, it also emphasizes the role of client contributions in the model’s development.
% Furthermore, \cite{RN206} enhance both ownership verification and traceability by employing a hybrid watermarking mechanism. This scheme integrates a global backdoor watermark with client-specific feature-based watermarks, leveraging continual learning to preserve model performance for both the primary task and watermark task.

\subsubsection{Client-side Watermarking}

\noindent Server-side watermarking typically assumes that the server can freely modify global model parameters to embed the watermark. Additionally, it presumes that the intellectual property of the global model belongs solely to the server, overlooking the users’ crucial contributions to model training.

A client-side backdoor watermarking scheme was developed to overcome the limitations of server-centric approaches and to recognize individual client contributions~\cite{RN209}. The integration of a gradient scaling mechanism enables this method to mitigate the challenge of limited client influence during watermark embedding.
% To address the above limitations of server-centric watermarking, \cite{RN209} proposed a client-side backdoor watermarking scheme that acknowledges individual client contributions.
% By introducing a gradient scaling mechanism, their approach mitigates the challenge of limited client influence on watermark embedding.
Building on this foundation, an improved client-side backdoor watermarking scheme~\cite{RN211} was proposed with a sophisticated trigger set generation strategy. This approach employs a permutation-based secret key and noise-based patterns to ensure non-ambiguous triggers and to prevent watermark forgery.
% Building on this foundation, \cite{RN211} introduced an improved client-side backdoor watermarking scheme with a more sophisticated trigger set generation method. Their non-ambiguous trigger set construction leverages a permutation-based secret key and noise-based patterns, preventing adversaries from forging watermarks.
%Building on Liu et al., Yang et al. \cite{RN211} introduced an improved client-side backdoor watermarking scheme with a more advanced trigger set generation method. They proposed a non-ambiguous trigger set construction method using a permutation-based secret key and noise-based patterns, effectively preventing adversaries from forging watermarks.
Similarly, an enhanced trigger set generation strategy was introduced to incorporate an additional private watermark class, thereby achieving watermark embedding without modifying the original dataset or degrading main task performance~\cite{chen2023client}.
% Similarly, \cite{chen2023client} proposed an enhanced trigger set generation method by incorporating an additional private watermark class, ensuring watermark embedding without altering the original dataset or degrading main task accuracy.
%Chen et al. \cite{chen2023client} proposed another advanced trigger set generation method for the client-side backdoor watermarking scheme. By introducing an additional private watermark class, they modified the model architecture from N-classification to N+1-classification during training without altering the original dataset or degrading main task accuracy.
FedIPR~\cite{Li2023FedIPR} extends FL to secure federated learning by integrating both feature-based and backdoor-based watermarking on the client side. 
%FedIPR \cite{Li2023FedIPR} further extends FL to a secure federated learning scenario by incorporating both feature-based and backdoor-based watermarking on the client side. It is compatible with advanced security techniques such as differential privacy, homomorphic encryption, and defensive aggregation, making it the first general framework in a secure FL setting.
To address watermark conflicts, FedMark~\cite{RN77} evaluates the capacity limitations of existing schemes and introduces a high-capacity watermarking approach using Bloom Filters.% This method enables conflict-free watermarking for a significantly larger number of participants, supporting approximately 150 clients, compared to the 40 clients of FedIPR.

%To address the issue of watermark conflicts, FedMark \cite{RN77} analyzes the capacities of existing watermarking schemes and introduces a large-capacity watermarking scheme that leverages Bloom Filters to enable conflict-free watermarking for a significantly larger number of participants—approximately 150, compared to FedIPR's 40.

\subsubsection{Dual-side Watermarking}

\noindent As both the server and clients contribute to model training in FL, it is essential for both parties to embed watermarks to assert ownership. 
To this end, a joint watermarking scheme~\cite{RN212} was proposed, where the server embeds backdoor-based watermarks while clients incorporate feature-based watermarks to enable mutual ownership verification.
% \cite{RN212} proposed a joint watermarking scheme, where the server embeds backdoor-based watermarks, while clients incorporate feature-based watermarks to enable mutual ownership verification.

Despite these advancements, SFL introduces unique challenges due to its asymmetric control paradigm. In SFL, clients lack direct access to most model parameters governed by the server, while the server has no control over clients’ raw data inputs or labels. This asymmetry necessitates a dual-side ownership verification mechanism, which is incompatible with existing watermarking schemes. Therefore, we next introduce a robust client-server watermarking scheme for SFL.

\begin{figure}[t]
    \centering
    \includegraphics[width=1.0\linewidth]{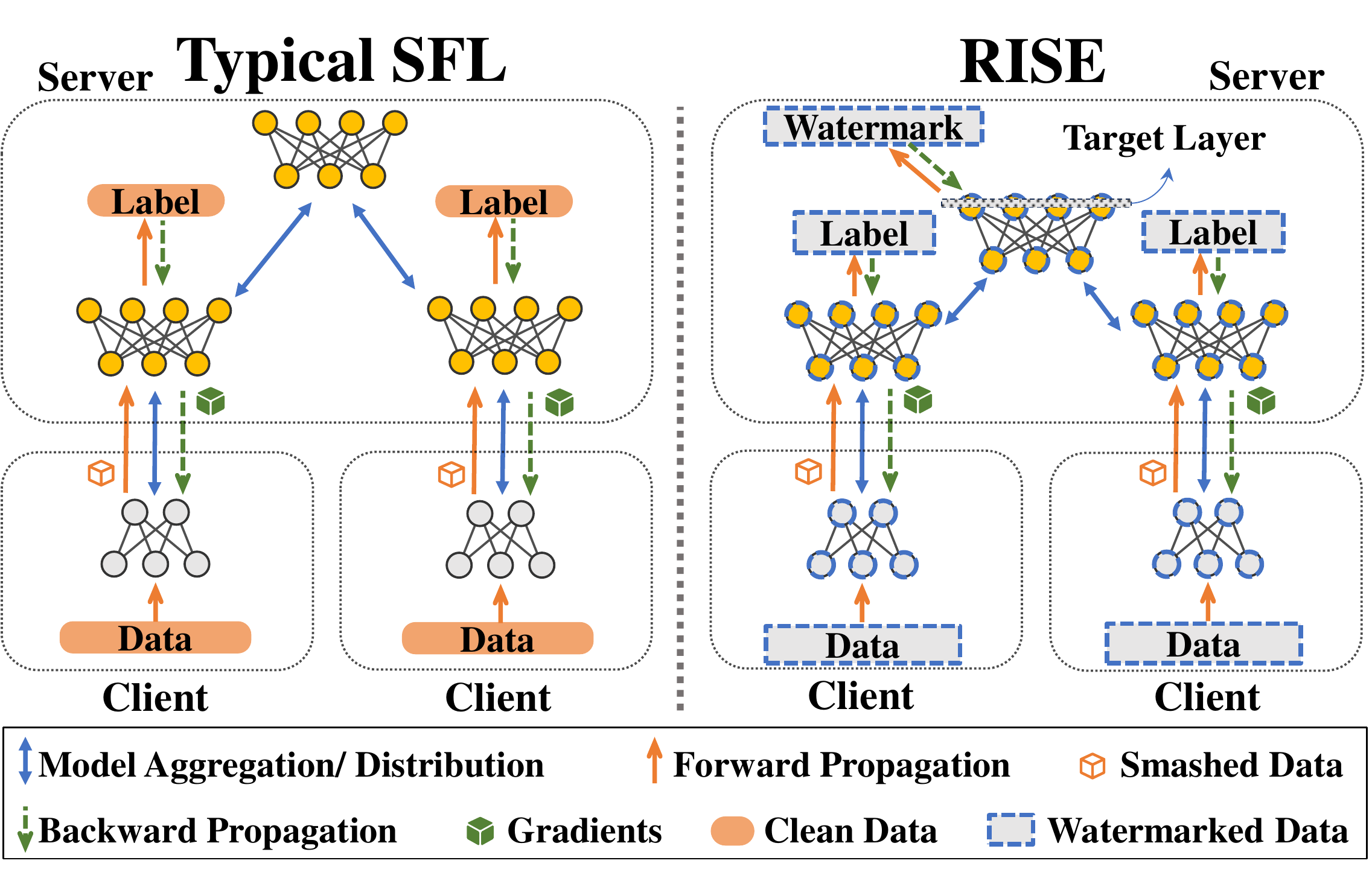}
    \caption{Typical SFL vs. RISE}
    \label{fig:sfl_framework}
\end{figure}

% \begin{table}[t]
% \centering
% \begin{tabular}{cl}
% \toprule
% \textbf{Notation} & \textbf{Description} \\
% \midrule
% $K$              & Number of clients in the SFL system \\
% $B_k$            & Bottom model of client $k$ \\
% $B_k^t$          & Bottom model of client $k$ at round $t$ \\
% $T^t$            & Top model at round $t$ \\
% $D_k$            & The dataset owned by client $k$ \\
% $n_k$            & Number of samples in $D_k$ \\
% $(x, y)$         & Sample-label pair \\
% $z$              & Smashed data \\
% $P_k$            & Trigger pattern of client $k$ \\
% $(x_k^*,y_k^*)$  & Backdoored data with trigger pattern $P_k$ \\
% $\rho$           & Ratio of backdoored samples in $D_k$ \\
% $\mathcal{T}_k$  & Client $k$'s backdoor dataset \\
% $\mathcal{L}^{t}_{k}$ & Main task loss of client $k$ at round $k$\\
% $L^{t}_{\text{WM}}$   & The loss of server-side watermarking at round $k$ \\
% $\alpha$         & The scale of $L^{t}_{WM}$ in overall loss function \\
% $Acc_{\text{main}}$   & Accuracy of main task \\
% $\theta_B$       & Detection rate of backdoor-based watermarking \\
% $\theta_F$       & Detection rate of feature-based watermarking \\
% $\sigma$         & The scale of differential privacy noise \\
% \bottomrule
% \end{tabular}
% \caption{Notations used in this article}
% \label{tab:notations}
% \end{table}

\section{Proposed Scheme}

% \noindent In this section, we explain the motivation, design rationale, detailed scheme and watermarking verification in sequence to offer a comprehensive view of RISE. 
\noindent In this section, we explain the design rationale, detailed scheme and watermarking verification to offer a comprehensive view of RISE. 

\subsection{Design Rationale and Detailed Scheme}

\noindent We now elaborate on the rationale for assigning distinct watermarking paradigms to the server and clients, demonstrating how this division naturally aligns with their respective roles within the SFL architecture. The right part of Fig.~\ref{fig:sfl_framework} shows an overview of RISE and Alg.~\ref{algo:overall_wm_embedding} presents the training process of RISE.

% 联合水印 算法流程图
\begin{figure}[t]
    \centering
    \includegraphics[width=1.0\linewidth]{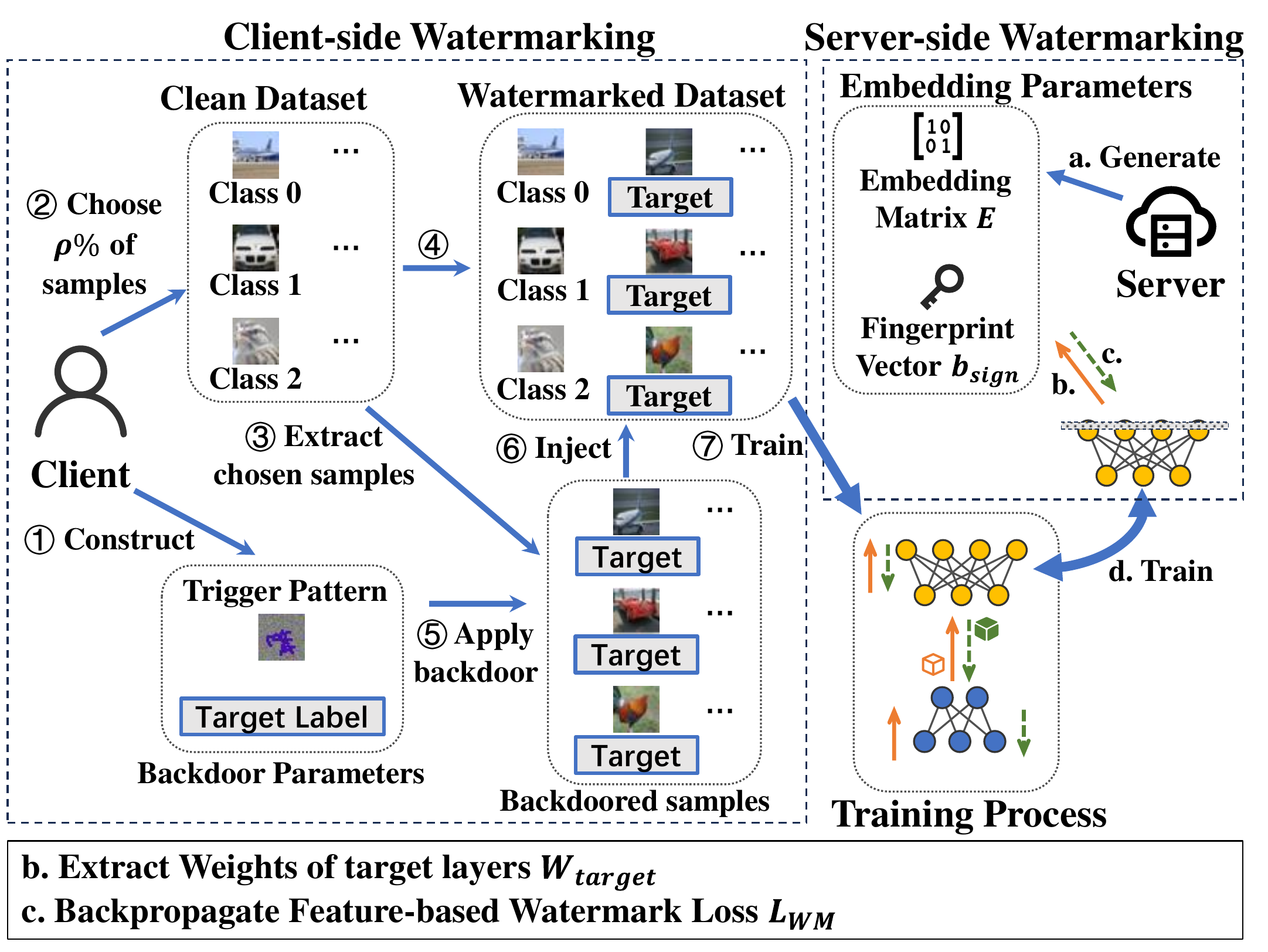}
    \caption{RISE Watermarking Scheme}
    \label{fig:joint_watermarking}
\end{figure}

\begin{algorithm}[h]
    \caption{ Training Process with Watermarking}
    \label{algo:overall_wm_embedding}
    \begin{algorithmic}[1]  % 开启手动行编号
        \Statex{\textbf{Server Initialization: }}
            \State Generate embedding matrix \(M\) and signature vector \(b_{sign}\) for server-side watermarking
            
        \Statex{\textbf{Client Initialization: }}
        \For{\(k\) in \(K\) clients}
            \State Generate trigger pattern \(P_k\) and choose target class \(y^*_K\)
            \For{\(\rho\) percent of \((x,y) \in D_k\)}
            \State Apply \(P_k\) to clean data \(x\) and modify the label \(y=y^*_k\)
            \EndFor
        \EndFor
        \For{each global round \(t\)}
        \State The server distributes global bottom model \(B^t\) of round \(t\) to each client
            \Statex \textbf{Client Training:}
            \For{\(k\) in \(K\) clients}
                \State Compute smashed data \(z_k=B^{t}_{k}(x)\), where \((x,y)\in D_k\)
                \State Send \(z_k\) and corresponding labels \(y\) to the server
                \State Get gradients \(g_k\)  from the server and backpropagate \(g_k\)
            \EndFor

            \Statex \textbf{Server Training:}
            \State Compute \(\hat{y}_k=T^{t}(z_k)\) for each client \(k\)'s \(z_k\)
            \State Compute loss \(\mathcal{L}^{t}_{k}\) by \(\hat{y}_k\)
            \State Backpropagate using \(\mathcal{L}^{t}_{k}\) and send gradients \(g_k\) to \(k\) when backpropagating to the split layer
            \State Compute feature-based watermark loss \(L^{t}_{\text{WM}}\) with the weights of target layers in \(T^t\).
            \State Backpropagate using \(\alpha L^{t}_{\text{WM}}\) and update \(T^t\)
            \State Aggregate bottom models \(\{B^{t}_{k}\}^{K}_{k=1}\) with FedAvg.
        \EndFor
    \end{algorithmic}
\end{algorithm}

% We now unpack the reasoning behind assigning different watermarking paradigms to the server and clients, respectively, and highlight how this division aligns with their roles in the SFL architecture.

\subsubsection{Feature-based Watermarking on the server side}

% \noindent Constructing a backdoor-based watermark on the server side requires synthesizing trigger patterns that resemble clients' smashed data \(z\). 
% While a generative adversarial network–based framework capable of generating such representations for backdoor attacks has been proposed \cite{Yu2024Chronic}, the approach relies on computationally intensive adversarial training and model inversion techniques.
% % While \cite{Yu2024Chronic} propose a generative adversarial network–based framework capable of generating such representations for backdoor attacks, their approach depends on computationally intensive adversarial training and model inversion techniques. 
% These requirements make backdoor-based watermarking impractical on the server side for our intellectual property protection scenario.

% In contrast, feature-based watermarking presents a more feasible alternative. Since the server controls the loss function and the backpropagation pipeline, it can embed a verifiable watermark into specific target layers without introducing intensive extra computing or significantly disrupting the primary training objective.

\noindent Constructing backdoor-based watermarks on the server side requires generating trigger patterns that mimic clients’ smashed data.
A recent work~\cite{Yu2024Chronic} demonstrated the feasibility of synthesizing such representations for backdoor injection, but it relies on adversarial training and model inversion, both of which incur substantial computational overhead.

In contrast, feature-based watermarking provides a more practical solution. 
Since the server controls the loss function and backpropagation pipeline, it can embed a verifiable watermark into target layers without additional heavy computation or disruption to the primary training objective.

%In contrast, feature-based watermarking offers a more tractable solution. As that the server controls the loss function and backpropagation pipeline, it can embed a verifiable watermark into the target layer(s) without introducing synthetic triggers or significantly interfering with the main training objective.

% Despite having full control over the bottom models during the FedAvg stage, the server does not embed any watermarks into the bottom models. 
% This is because the FedAvg service for bottom models is a relatively trivial contribution compared with the overall training process and there is no need to claim the contribution by watermarks. 
% Furthermore, the bottom model is too small to carry significant additional information (the bottom model is only approximately 360KB, compared to 42MB for the top model in our ResNet18-based SFL implementation), the only mission for them is to extract characteristics from clients' data.

Thus, an auxiliary term is incorporated alongside the primary task loss to embed the feature-based watermark. Let \(L_{\text{WM}}\) denote the server-side watermarking loss, computed over secretly selected layer(s) using a private signature vector \(b_{sign}\) and a secret embedding matrix \(M\). The overall loss \(L\) can be formulated as:
\begin{equation}
    L=\mathcal{L}+\alpha L_{\text{WM}},
\end{equation}
where \(\mathcal{L}\) is the main task described in Eq.~(\ref{eq:main_task_formalization}), and \(\alpha\) controls the strength of the watermarking regularization term.

A series of prior works have demonstrated that embedding feature-based watermarks into batch normalization (BN) layers can achieve both stealthiness and robustness against classic watermark removal attacks~\cite{Li2023FedIPR}. Accordingly, in RISE, the server-side watermark is embedded in BN layers in the top model to protect the server's intellectual property. To achieve this, the weights of the BN layers $W_{BN}$ are optimized according to Eq.~(\ref{eq:uchida_wm}).
\begin{equation}
\label{eq:uchida_wm}
H(W_{\text{BN}} \cdot M) = b_{sign},
\end{equation}where $M$ is a randomly generated embedding matrix, $b_{\text{sign}} = \{0 , \, 1\}^{N}$ is the signature vector to be embedded into $W_{\text{BN}}$, both kept secret and $H(x)$ is the Heaviside function:
\begin{equation}
H(x)=\left\{
\begin{array}{l}
1  \;\;\;x \ge 0, \\
0 \;\;\; else.
\end{array}
\right.
\end{equation}

% Uchida et al. \cite{Uchida2017-ds} convert Eq.~\ref{eq:uchida_wm} into the initial embedding loss function $\mathcal{E}(W)$ by using binary cross entropy described as below:

% \begin{equation}
% \mathcal{E}(W)=-(b_{sign} \ log(y) + (1-b_{sign}) \  log(1 - y))
% \end{equation}

% where $y=\sigma(W \cdot M)$ ($\sigma$ is the sigmoid function).

In the proposed watermarking scheme illustrated in Fig.~\ref{fig:joint_watermarking}, the server-side feature-based watermark is embedded through the following regularization term:
\begin{equation}
\label{eq:loss_feature}
\begin{split}
L_{\text{WM}}(W_{\text{BN}}, M, b_{\text{sign}})&=\sum^{N}_{i=1} Hinge(b_i \cdot y_i)
\\
&=\sum^{N}_{i=1}max(1-b_i \cdot y_i,\ 0),
\end{split}
\end{equation}
where \(b_i\) denotes the \(i\)-th bit of the binary signature vector \(b_{\text{sign}}\), 
and \(y_i\) is the watermark-related response extracted from the BN layer via embedding matrix \(M\), defined as
\begin{equation}
\label{eq:y_i_in_feature_watermark}
y_i = W_i \cdot M_i,
\end{equation}
where \(W_i\) is its \(i\)-th row of \(W_{\text{BN}}\) and \(M_i\) is the \(i\)-th column of \(M\). 
The Hinge-like formulation \(Hinge(\cdot)\) encourages the sign of each response \(y_i\) to align with the corresponding bit \(b_i\), 
thereby embedding a verifiable binary watermark into the model parameters.

\subsubsection{Backdoor-based Watermarking on the client side}

% \noindent In feature-based watermarking schemes of FL, clients complete the entire training process locally, and have full control over both the model architecture and the loss function, allowing them to embed watermarks by introducing additional regularization terms specifically designed for watermarking purposes.
% In contrast, SFL clients are limited to forward-only computation using the bottom model, while both loss computation and optimization are carried out by the server. This architectural separation makes feature-based watermarking infeasible on the client side, as it requires access to the loss function and participation in the full optimization pipeline.
\noindent In FL, each client performs full local training and can modify both the model architecture and loss function, enabling feature-based watermark embedding through additional regularization terms. 
However, in SFL, clients are restricted to forward propagation using the bottom model, while loss computation and parameter optimization are entirely handled by the server. 
This limited control prevents clients from accessing the loss function or participating in backpropagation, thereby rendering feature-based watermarking infeasible on the client side.

%In conventional feature-based watermarking schemes, clients complete the entire training process locally before submitting their trained models for global aggregation. In such settings, clients have full control over both the model and the loss function, enabling them to embed watermarks by incorporating additional regularization terms tailored to the watermarking task.

%In contrast, SFL clients are restricted to forward-only computation with the bottom model, whereas both loss computation and optimization are performed by the server. This architectural separation renders feature-based watermarking infeasible on the client side, as it requires access to the loss function and full participation in the optimization process.  

% As a result, injecting backdoored samples into the local dataset emerges as the most practical approach for client-side watermarking. During training, the effects of these backdoor-based watermarks are naturally propagated from the bottom model to the top model via standard forward and backward passes.

% This mechanism enables clients to later verify their ownership by querying the trained model with poisoned samples containing their private trigger patterns. Successful activation of the backdoor indicates that the client's private data and, therefore, their contribution to the training process have been embedded in the SFL model in a verifiable way.

As a result, injecting backdoored samples into local datasets becomes the most practical approach for client-side watermarking. 
During training, the watermarking is naturally propagated from the bottom model to the top model through standard forward and backward passes, enabling clients to verify ownership by later querying the trained model with their private trigger patterns. 
A successful activation of the backdoor confirms that the client’s private data--—and thus their contribution--—has been embedded in the SFL model in a verifiable manner.

Under these constraints, triggers must be carefully designed to satisfy three key principles:

\begin{enumerate}
    \item \textbf{Stealthiness}: Triggers should blend naturally with normal data, resisting standard anomaly detection methods. 
    \item \textbf{Distinctiveness}: Triggers must be sufficiently learnable while minimizing negative impact on main task accuracy. 
    \item \textbf{Robustness}: Watermarks should resist common removal techniques, safeguarding clients' rights over models.
\end{enumerate}

% \begin{enumerate}
%     \item \textbf{Stealthiness}: Triggers should blend naturally with benign data, remaining undetectable by standard anomaly detection methods. 
%     \item \textbf{Distinctiveness}: Triggers must be sufficiently learnable to ensure reliable activation while minimizing degradation of main-task accuracy. 
%     \item \textbf{Robustness}: Watermarks should withstand common removal techniques, safeguarding clients’ ownership over the model.
% \end{enumerate}

% As shown in Fig.~\ref{fig:joint_watermarking}, the proposed backdoor-based watermarking scheme, following a data-driven approach, randomly selects a small proportion of clean data and relabels them to a target class , which is privately selected from the original label space. 
% Client \(k\)'s private trigger pattern $P_k$ is then applied to these samples, following the aforementioned key principles. 
% Subsequently, the backdoored samples are incorporated into the clean data, forming the client-specific watermarked dataset \(D_k\), which is used in the standard training process. These samples are designed to induce a specific misclassification behavior that can be reliably and uniquely triggered by the client’s private backdoor, thereby enabling verifiable ownership attribution.

As illustrated in Fig.~\ref{fig:joint_watermarking}, the proposed backdoor-based watermarking scheme first initializes clients' trigger patterns and randomly selects a small portion of clean samples, relabeling them to a privately chosen target class from the original label space. 
Each client \(k\) then applies its private trigger pattern \(P_k\) to these samples in accordance with the above principles. 
The resulting backdoored samples are merged with the clean dataset to form the client-specific watermarked dataset \(D_k\), which is used in standard training. 
These samples induce a deterministic misclassification that can be uniquely activated by the client’s private trigger, thereby enabling verifiable ownership attribution. Consequently, the client-side watermarking objective can be formulated as:
\begin{equation}
    T(B_{k}(x^*))=y^* \;\; for (x^*,y^*)\in \mathcal{T}_k ,
\end{equation}
where \(T(\cdot)\) and \(B_{k}(\cdot)\) represent the outputs of top model \(T\) and bottom model \(B_{k}\) respectively. \((x^*,y^*)\) represents the backdoored data-label pair in client \(k\)'s backdoor dataset \(\mathcal{T}_k\).

As training progresses, backdoored samples propagate through SFL models and blend into the learned feature space.

\subsection{Ownership Verification} 
\label{sec:ownership_verification}

\noindent In this section, we illustrate the ownership verification processes for both server-side and client-side watermarking.

\subsubsection{Server-side Ownership Verification} 

\begin{algorithm}[htbp]
    \caption{ Verification process for top model}
    \label{algo:ownership_verification_top}
    \begin{algorithmic}[1]  % 开启手动行编号
        \Require Embedding Matrix \(M\);
        Signature Vector \(b_{sign}\);
        Suspicious Bottom Model \( T_{sus} \)
        \Ensure Feature-based watermarking detection rate \(\theta_F\)
        \State Extract the weights \(W\) from the target layer of suspicious model \(T_{sus}\)
        \State Calculate \(y = W \cdot M\)
        \State Calculate \(\theta_{F} = 1 - \frac{1}{N}\sum_{i=1}^{N}{H(- b_{i} \cdot y_{i})}\)
        \State return \(\theta_F\)
    \end{algorithmic}
\end{algorithm}

\noindent The robustness of the feature-based watermark embedded in the top model can be directly assessed using the secret embedding matrix $M$ and the signature vector $b_{sign}$ by computing the loss over the weights of the target layers, following the same formulation as Eq.~(\ref{eq:loss_feature}). However, the magnitude of the loss is not straightforward to interpret. To provide a more intuitive and interpretable evaluation, we introduce a percentage-based metric, described as Eq.~(\ref{eq:theta_f}).
\begin{equation}
\label{eq:theta_f}
\theta_{F} = 1 - \frac{1}{N}\sum_{i=1}^{N}{H(- b_{i} \cdot y_{i})},
\end{equation}
where $y_i$ is defined in Eq.~(\ref{eq:y_i_in_feature_watermark}) and $N$ is the bit length of $b_{\text{sign}}$. According to the optimization objective in Eq.~(\ref{eq:loss_feature}), the $i$-th bit of the watermark is successfully embedded when \(H(- b_{i} \cdot y_{i}) = 0\). Therefore, \(\theta_{F}\) represents the ratio of successfully embedded bits of the feature-based watermark. 
Server-side ownership verification procedure is outlined in Alg.~\ref{algo:ownership_verification_top}.

\subsubsection{Client-side Ownership Verification} 

\noindent Given the constraints of backdoor-based watermarking, verification should be performed stealthily, similar to the embedding process. This involves incorporating backdoored samples into the normal test samples. 

Specifically, client $k$ computes the backdoor-based watermark detection rate $\theta_B^k$ by evaluating the server's responses to the smashed data generated from the backdoored dataset $\mathcal{T}_k$. If $\theta_B^k$ exceeds a predefined threshold $\tau$, the backdoor-based watermarking of client $k$ is considered successfully embedded into both the bottom and top models. 
The verification process for the suspicious bottom model is summarized in Alg.~\ref{algo:ownership_verification_bottom}.

\begin{algorithm}[htbp]
    \caption{Verification process for bottom model}
    \label{algo:ownership_verification_bottom}
    \begin{algorithmic}[1]  % 开启手动行编号
        \Require Clean Test Dataset \( \mathbb{D} \);
        Trigger Pattern \( P_k \); Backdoor Ratio \( \rho \); Target Class \( Y^k_t \);
        Suspicious Bottom Model \( B_{sus} \)
        \Ensure Backdoor-based watermarking detection rate \(\theta_B\)
        
        \State Randomly apply \(P_k\) to \( \rho \) fraction of \( \mathbb{D} \)
        \State Send \(z=B_{sus}(\mathbb{D})\) to the server 
        \State The server returns \(y=T(z)\) to client \(k\)
        \State Match backdoored samples among \(y\) with target label \(Y^k_t\)
        \State Calculate the detection rate \(\theta_B\)
        \State return \(\theta_B\)
    \end{algorithmic}
\end{algorithm}

Since backdoor-based watermarks propagate throughout both SFL models, clients can verify the ownership of both the bottom and top models. Furthermore, if an adversary steals and combines both side models into a new full model, clients can still assert ownership by triggering the backdoors exclusively owned by them.

\section{Experiments}

\noindent In this section, we present our empirical study on RISE, evaluating the significance, fidelity and watermark robustness under various settings.
%Experimental results demonstrate that RISE provides a robust and effective solution for intellectual property protection in split federated learning. 
% Our source codes for implementation are available at \href{https://github.com/dickton/sfl_watermarking/}{\includegraphics[height=2.5ex]{images/github-mark.png}}.

\subsection{Experimental Settings}
\label{main_experimental_setting}
We elaborate the main experimental settings.

\textbf{Baseline.} We use clean SFL and our implementation of FedIPR~\cite{Li2023FedIPR} in SFL trained under identical settings for comparison.

\textbf{Dataset.} We evaluate RISE on standard datasets: CIFAR-10 (C10)~\cite{krizhevsky2009learning}, CIFAR-100 (C100)~\cite{krizhevsky2009learning}, Tiny-ImageNet (TINet)~\cite{le2015tiny} and Fashion-MNIST (FM)~\cite{xiao2017fashion}.

\textbf{Split Scheme.} Given that the core idea of SFL is the trade-off between computation offloading and transmission expenses, we split ResNet-18~\cite{He2016ResNet18} after the first Residual Block, following~\cite{thapa2022splitfed}. MobileNetV2~\cite{Sandler2018MobileNetV2} and DenseNet-121~\cite{Huang2017DenseNet121} are split at the split point 2 (see Suppl.~\ref{sec:split_scheme_for_mobilenet} and Suppl.~\ref{sec:split_scheme_for_densenet}).
Unless otherwise specified, this split configuration is used throughout experiments. 
% For more detailed settings related to the splitting scheme, please refer to

\textbf{Hardware and software Environment. }The experiments were conducted in a hardware and software environment, as summarized in Table~\ref{tab:env_config}.

\begin{table}[htbp]
\centering
\begin{tabular}{ll}
\toprule
\textbf{Component} & \textbf{Specification} \\
\midrule
GPU & NVIDIA RTX 5090 \\
CPU & 24 vCPU Intel Core i9-14900K \\
Memory & 128GB \\
Operating System & Ubuntu 22.04.4 LTS \\
CUDA Version & 12.8 (V12.8.61) \\
Python Version & 3.12.11 \\
\bottomrule
\end{tabular}
\caption{Experimental Environment Configuration}
\label{tab:env_config}
\end{table}

\textbf{Other Settings.} Default number of clients is 10 unless otherwise specified. For more experimental settings, please refer to Suppl.~\ref{sec:more_exp_detail}.

\subsection{Watermark Statistical Significance}

\noindent To evaluate the effectiveness and statistical significance of embedded watermarks, we report the detection rates for both client-side and server-side watermarking, denoted as \(\theta_B\) and \(\theta_F\), respectively, along with their corresponding $p$-values, $p_B$ and $p_F$ computed using the Mann-Whitney U test. Experiments are conducted on CIFAR-10 under three representative numbers of clients.

\begin{table}[t]
    \centering
    \begin{tabular}{lcccc}
        \toprule
        \textbf{\(K\)}   & \textbf{\(\theta_B\)} & \textbf{$p_B$}    & \textbf{\(\theta_F\)} & \textbf{$p_F$}    \\
        \midrule
        10          & \(99.32\%\)    & 0.011925  & \(100.00\%\)   & 0.007495    \\
        20          & \(99.64\%\)    & 0.007937  & \(100.00\%\)   & 0.007089    \\
        50          & \(94.18\%\)    & 0.007937   & \(100.00\%\)  & 0.005584    \\
        \bottomrule
    \end{tabular}
    \caption{\textnormal{Table shows client-side watermarking detection rates \(\theta_B\) and server-side watermarking detection rates \(\theta_F\) under different numbers of clients \(K\) with corresponding $p$-values.}}
    \label{tab:varying_clients_wm_significance}
\end{table}

As shown in Tab.~\ref{tab:varying_clients_wm_significance}, both sides of watermarking achieve high detection rates, with \(\theta_\text{B}\) consistently above \(94\%\) and \(\theta_\text{F}\) maintained at \(100\%\). All corresponding \(p\)-values fall below 0.012, confirming the statistical significance of observed watermarking effects. The results demonstrate the effectiveness and statistical significance of the RISE scheme, providing strong evidence to support ownership claims.

% \begin{table*}[h]
%     \centering
%     \begin{tabular}{c ccccccccccc}
%                 \toprule
%                 \multirow{2}{*}{\textbf{\(K\)}} & \multicolumn{3}{c}{$Acc_{\text{main}}$(\%)} & \multirow{2}{*}{\(p^{\text{C}}_{\text{main}}\)} & \multirow{2}{*}{\(p^{\text{S}}_{\text{main}}\)} & \multicolumn{2}{c}{$\theta_{\text{B}}$(\%)} & \multirow{2}{*}{\(p_{\text{B}}\)} & \multicolumn{2}{c}{$\theta_{\text{F}}$(\%)} & \multirow{2}{*}{\(p_{\text{F}}\)}\\
%                 \cmidrule(lr){2-4} \cmidrule(lr){7-8} \cmidrule(lr){10-11}
%                 & RISE & Client-only & Server-only& & &  RISE & Client-only & & RISE& Server-only \\
%                 \midrule
%                 10  & 79.30 & \textbf{81.58} &  &  &  & \textbf{100.00} & 99.32 &  \\
%                 20  & 76.62 & \textbf{79.36} &  &  &  &  99.43 & \textbf{99.63}  &  \\
%                 30  & 74.41 & \textbf{77.85} &  &  &  &  89.29 & \textbf{98.97}  &  \\
%                 40  & 73.34 & \textbf{75.18} &  &  &  &  58.20 & \textbf{98.22}  &  \\
%                 50  & 72.67 & \textbf{74.55} &  &  &  &  67.06 & \textbf{93.87}  &  \\
%                 \bottomrule
%             \end{tabular}
%     \caption{\textnormal{Table shows client-side watermarking detection rates \(\theta_B\) and server-side watermarking detection rates \(\theta_F\) under different numbers of clients \(K\) with corresponding $p$-values.}}
%     \label{tab:mutual_inference_grand_table}
% \end{table*}

\subsection{Ablation Study on Watermarking}
% \noindent We conducted mutual interference experiments to explore the potential conflicts caused by watermarking process of RISE. To evaluate potential conflicts, we conducted a series of experiments testing the main task performance \(Acc_{\textnormal{main}}\) among vanilla SFL without any watermarking (denoted as \textbf{Clean}), RISE, SFL with only client-side watermarking of RISE (\textbf{$C_{\text{only}}$}) and SFL with only server-side watermarking of RISE (\textbf{$S_{\text{only}}$}), all of which are trained with the identical hyperparameters. We furthered the experiment by training RISE with backdoored samples proportion \(\rho\) varying. All of the aforementioned experiments are performed with 10 clients on the CIFAR-10 dataset.

\noindent We conducted mutual interference experiments to examine whether the client- and server-side watermarks interfere with each other.
Specifically, we compared the main-task accuracy among vanilla SFL without watermarking (\textbf{Clean}), RISE, RISE with only client-side watermarking (\textbf{$C_{\text{only}}$}), and RISE with only server-side watermarking (\textbf{$S_{\text{only}}$}), all trained under identical hyperparameters. 
% We further extended the experiments by varying the proportion of backdoored samples \(\rho\). 
Unless otherwise specified, experiments were performed on CIFAR-10.

\begin{table}[b]
    \centering
    \begin{tabular}{lcccc}
        \toprule
        \textbf{\(K\)}   & \textbf{Clean} & \textbf{RISE}    & \textbf{$C_{\text{only}}$} & \textbf{$S_{\text{only}}$}    \\
        \midrule
        10          & \(81.85\%\)    & \(81.78\%\)  & \(82.66\%\)   & \(82.01\%\)    \\
        20          & \(79.85\%\)    & \(79.36\%\)  & \(81.41\%\)   & \(80.87\%\)    \\
        30          & \(77.88\%\)    & \(77.85\%\)  & \(80.48\%\)   & \(78.03\%\)    \\
        40          & \(76.21\%\)    & \(75.18\%\)  & \(79.69\%\)   & \(77.62\%\)    \\
        50          & \(75.29\%\)    & \(74.55\%\)  & \(78.75\%\)   & \(77.15\%\)    \\
        \bottomrule
    \end{tabular}
    % \caption{\textnormal{Table shows the main task accuracy of clean SFL, RISE, SFL with only client-side watermarking of RISE and SFL with only sever-side watermarking of RISE on the CIFAR-10 with varying client numbers.}}
    \caption{\textnormal{Table shows the main-task accuracy of Clean SFL, RISE, single-side-only RISE on CIFAR-10 in different client numbers.}}
    \label{tab:mutual_inference}
\end{table}

\begin{table}[t]
    \centering
    \resizebox{\linewidth}{!}{
        \begin{tabular}{ccccccc}
            \toprule
            \multirow{2}{*}{\textbf{\(K\)}}   & \multicolumn{2}{c}{$\theta_{\text{B}}$(\%)} & \multirow{2}{*}{\(p_{\text{B}}\)} & \multicolumn{2}{c}{$\theta_{\text{F}}$(\%)} & \multirow{2}{*}{\(p_{\text{F}}\)}\\
            \cmidrule(lr){2-3} \cmidrule(lr){5-6}
            & RISE & \textbf{$C_{\text{only}}$} &  & RISE & \textbf{$S_{\text{only}}$} & \\
            \midrule
            10  & 99.32  & 99.86  & 0.1368 &  100.00  & 100.00  & 0.5000 \\
            20  & 99.63  & 99.38  & 0.8834 &  100.00  & 100.00  & 0.5000 \\
            30  & 98.97  & 97.17  & 0.9113 &  100.00  & 100.00  & 0.5000 \\
            40  & 98.22  & 98.93  & 0.0996 &  100.00  & 100.00  & 0.5000 \\
            50  & 93.87  & 91.73  & 0.7085 &  100.00  & 100.00  & 0.5000 \\
            \bottomrule
        \end{tabular}
    }
    \caption{\textnormal{Watermarking detection rates of RISE and single-side RISE \(K\) with corresponding $p$-values in different client number.}}
    \label{tab:mutual_inference_grand_table}
\end{table}

As presented in Tab.~\ref{tab:mutual_inference}, the difference of the main task accuracy \(Acc_{\textnormal{main}}\) among the 4 training modes is minor. To explore the potential mutual inference between client-side and server-side watermarks, we conducted watermarking efficiency tests of RISE and each single-side watermarking scheme. 
As shown in Tab.~\ref{tab:mutual_inference_grand_table}, the $p$-value of $\theta_{\text{B}}$ between RISE and $C_{\text{only}}$ is not significant, indicating that adding server-side watermarking does not interfere with the detection of client-side watermarks.
Similarly, the $p$-value of $\theta_{\text{F}}$ between RISE and $S_{\text{only}}$ shows no significant difference, suggesting that the presence of client-side watermarks does not affect server-side detection.

% \begin{table}[htbp]
% \centering

% \begin{tabular}{ccccc}
% \toprule
% \textbf{Split} & 
% $\boldsymbol{Acc_{\text{main}}}$ (\%) & 
% $\boldsymbol{\theta_{\text{B}}}$ (\%)& 
% $\boldsymbol{\theta_{\text{F}}}$ (\%)& 
% $\boldsymbol{\phi}$ \\
% \midrule
% 1    & 81.65 & 99.70  & 100.00 & 2.440 \\
% 2    & 81.78 & 99.32  & 100.00 & 2.320 \\
% 3    & 82.93 & 100.00 & 100.00 & 1.978 \\
% 4    & 83.12 & 100.00 & 100.00 & 1.601 \\
% 5    & 83.60 & 99.90  & 100.00 & 1.405 \\
% \bottomrule
% \end{tabular}
% \caption{Influence of different split points on task accuracy, watermark detection, and time cost ratio.}
% \label{tab:split_ratio_results_cifar10}
% \end{table}

\begin{table}[b]
\centering
\resizebox{\linewidth}{!}{
\begin{tabular}{cccccc}
\toprule
\textbf{Model} & \textbf{Split} & 
$\boldsymbol{Acc}$ (\%)& 
$\boldsymbol{\theta_{\text{B}}}$ (\%)& 
$\boldsymbol{\theta_{\text{F}}}$ (\%)& 
$\boldsymbol{\phi}$ \\
\midrule
\multirow{3}{*}{RN18}
  & 1    & 81.65 & 99.70  & 100.00 & 2.440 \\
  & 2    & 81.78 & 99.32  & 100.00 & 2.320 \\
  & 3    & 82.93 & 100.00 & 100.00 & 1.978 \\
\midrule
\multirow{3}{*}{MNV2}
  & 1        & 85.04 & 100.00 & 100.00 & 10.16\\
  & 2        & 84.99 & 99.97  & 100.00 & 8.108 \\
  & 3        & 85.14 & 99.90  & 100.00 & 3.153 \\
\midrule
\multirow{3}{*}{DN121}
  & 1        & 81.99 & 98.80  & 100.00 & 9.249 \\
  & 2        & 83.07 & 99.90  & 100.00 & 6.800 \\
  & 3        & 84.33 & 100.00 & 100.00 & 3.366 \\
\bottomrule
\end{tabular}
}
\caption{Comparison of different split points and models on task accuracy, watermark detection, and time cost ratio.}
\label{tab:split_model_results}
\end{table}

% The proportion of backdoored samples in the clients' private dataset \(D_k\) shows no noticeable impact on the main task performance according to the results in Tab.~\ref{tab:mutual_inference_varying_rho}.  

Thus, there is no clear evidence of interference between the client- and server-side watermarks, supporting the compatibility of RISE’s dual-side design. 

Furthermore, we investigate whether different splitting schemes in SFL affect the consistency of RISE. Tab.~\ref{tab:split_model_results} reports the main metrics of RISE under various splitting settings, where \(\phi\) denotes the server-to-client computation time ratio. See Suppl.~\ref{sec:detailed_split_scheme} for a more detailed ablation study on split schemes. Results show negligible differences across splits, indicating that RISE remains stable regardless of the splitting configuration.

% \begin{table}[b]
%     \centering
%     \begin{tabular}{lcccccc}
%         \toprule
%         \(\rho\)    & \(0.05\) & \(0.10\)    & \(0.15\)  & \(0.20\)   & \(0.25\)   & \(0.30\) \\
%         \midrule
%         \(Acc_{\textnormal{main}} \  (\%)\)     & \(81.63\)    & \(81.53\)  & \(81.27\)   & \(81.02\) &  \(81.20\)  &  \(80.77\) \\
%         \(\theta_{\textnormal{B}} \  (\%)\)     & \(99.38\)    & \(99.70\)  & \(99.93\)   & \(99.70\) &  \(99.90\)  &  \(99.93\) \\
%         \bottomrule
%     \end{tabular}
%     \caption{\textnormal{Table shows main task accuracy and client-side watermarking detection rate with the proportion of backdoored sample \(\rho\) varying.}}
%     \label{tab:mutual_inference_varying_rho}
% \end{table}

\subsection{Fidelity}
% \noindent In this section, we evaluate the fidelity of the proposed RISE scheme using both overall task metrics and its robustness under Neural Cleanse \cite{Wang2019NC}.
% \noindent This section evaluates RISE in terms of main-task fidelity and robustness against Neural Cleanse~\cite{Wang2019NC}.

% \subsubsection{Fidelity in Overall Metrics}
% \noindent To assess the fidelity of RISE, we compare its main task performance \(Acc_{\text{main}}\) and the average epoch time per training round with those of the standard SFL on CIFAR-10 and Fashion-MNIST datasets, across varying numbers of clients ranging from 10 to 50.
\noindent To evaluate the fidelity of RISE, we compare its main-task accuracy and average training time per epoch with those of clean SFL.

% 主任务箱线图
\begin{figure}[t]
    \centering
    \begin{subfigure}[t]{0.49\linewidth} % 使用 [t] 选项可以帮助顶部对齐
        \centering
        \includegraphics[width=\linewidth]{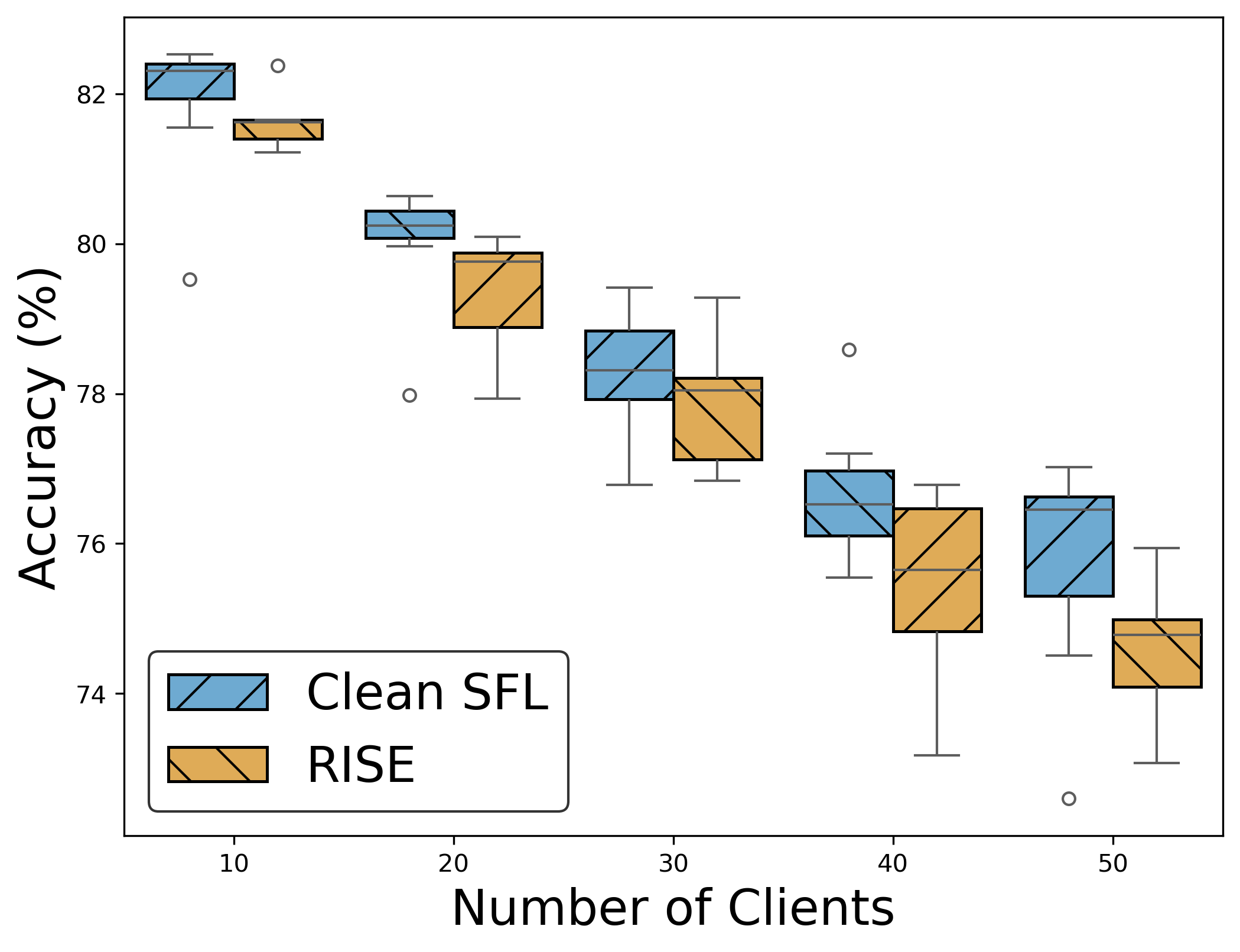}
        \caption{\textnormal{CIFAR-10}}
        \label{fig:cifar_10_acc_box_plot}
    \end{subfigure}% <--- 在这里加上 % 号，这是关键！
    \hfill
    \begin{subfigure}[t]{0.49\linewidth} % 使用 [t] 选项可以帮助顶部对齐
        \centering
        \includegraphics[width=\linewidth]{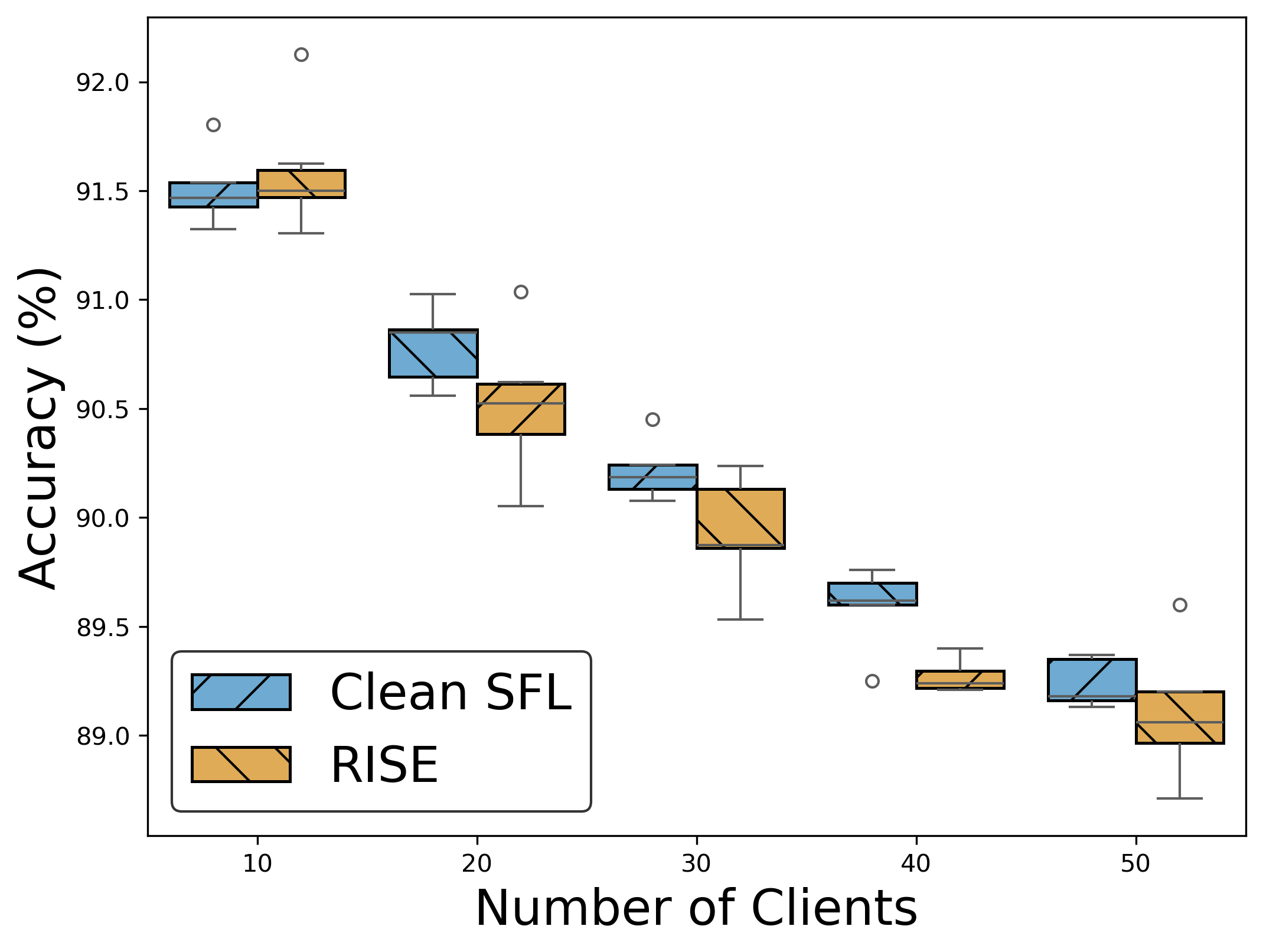}
        \caption{\textnormal{Fashion-MNIST}}
        \label{fig:cifar_10_running_time_box_plot}
    \end{subfigure}
    \caption{\textnormal{Figure presents \(Acc_{\text{main}}\) of clean SFL and RISE as client number varies with CIFAR-10 and Fashion-MNIST.}}
    \label{fig:fidelity_main_task}
\end{figure}

% 运行时间箱线图
\begin{figure}[t]
    \centering
    \begin{subfigure}[]{0.49\linewidth}
        \centering
        \includegraphics[width=\linewidth]{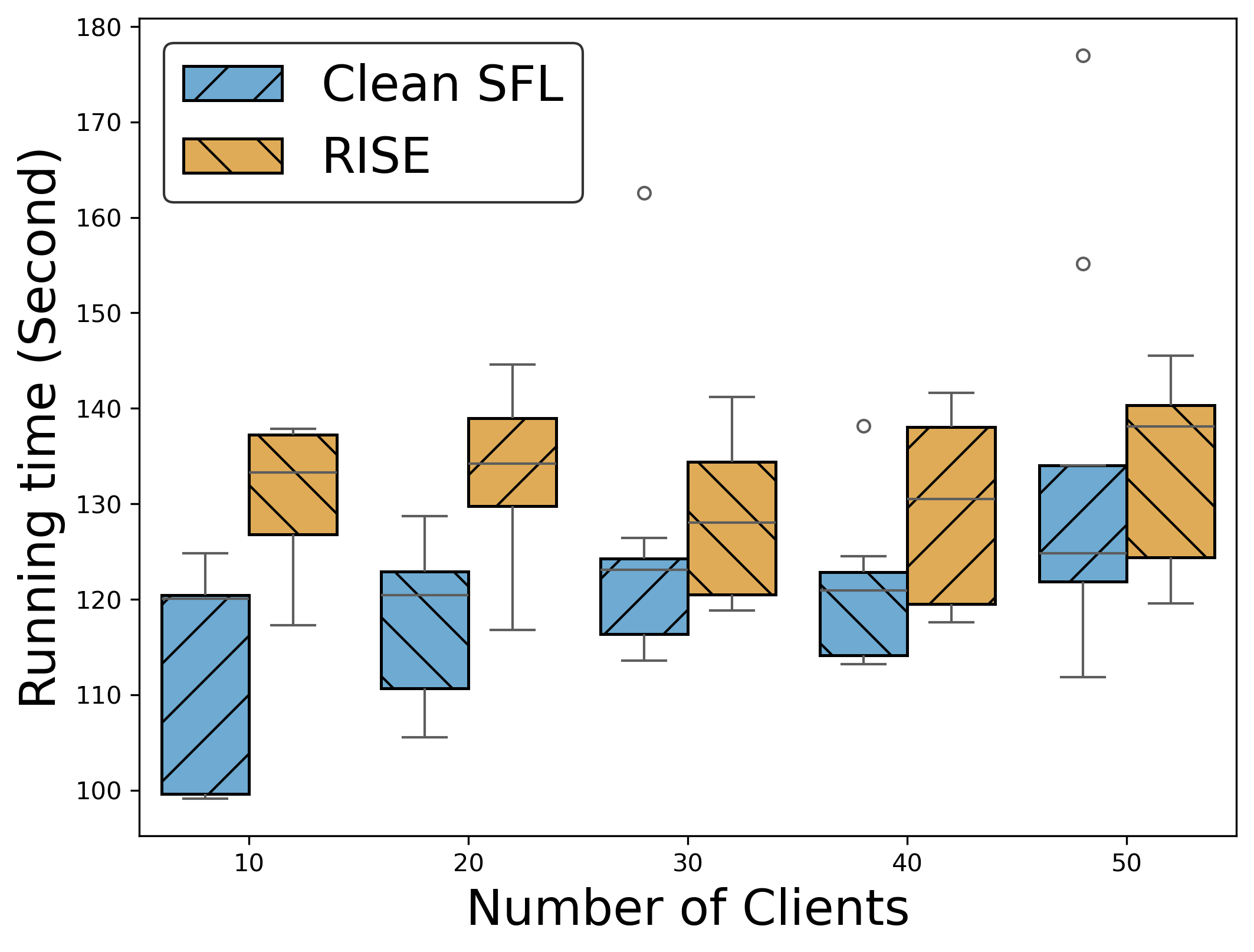}
        \caption{\textnormal{CIFAR-10}}
        \label{fig:cifar_10_acc_box_plot}
    \end{subfigure}
    \hfill
    \begin{subfigure}[]{0.49\linewidth}
        \centering
        \includegraphics[width=\linewidth]{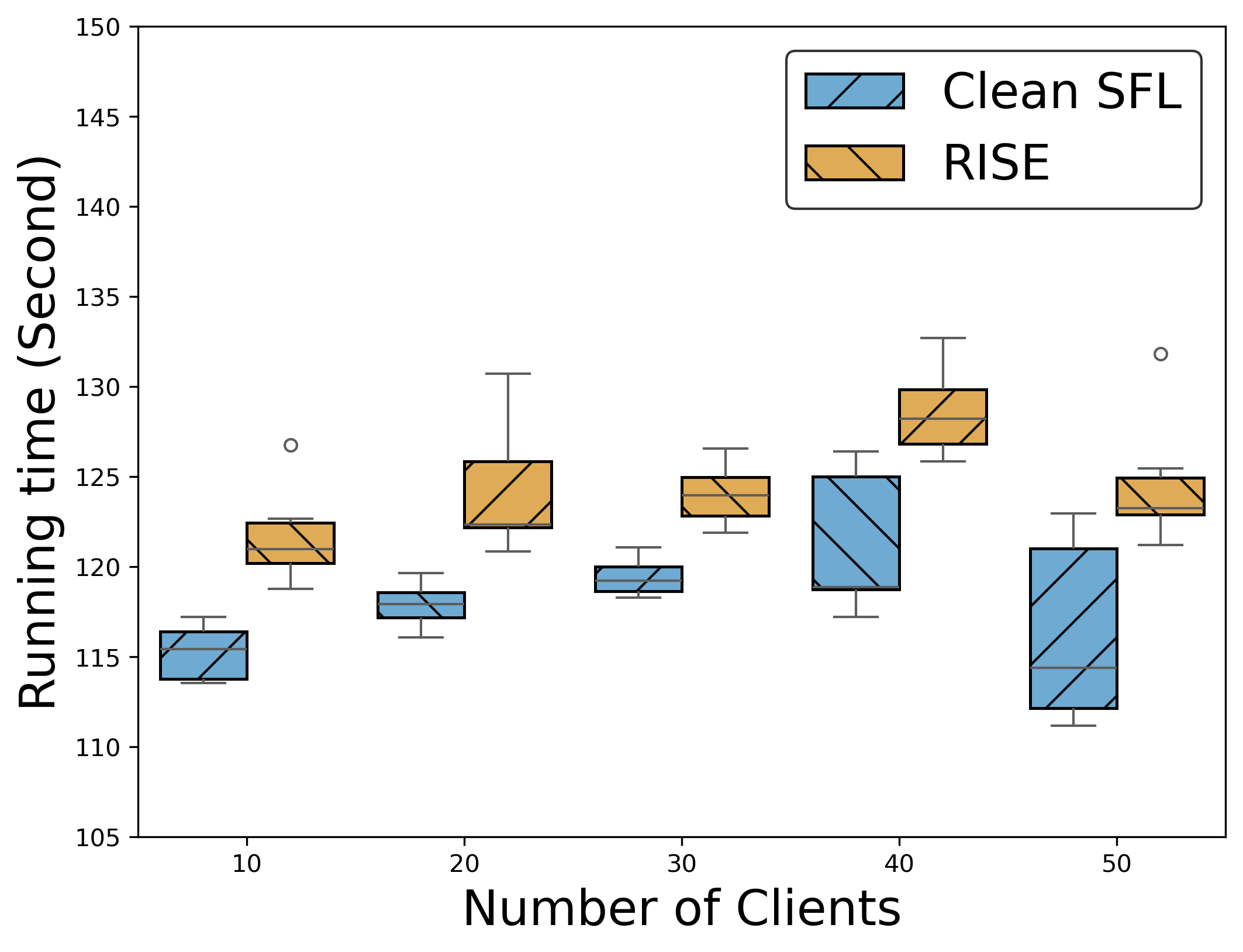}
        \caption{\textnormal{Fashion-MNIST}}
        \label{fig:cifar_10_running_time_box_plot}
    \end{subfigure}
    \caption{\textnormal{Figures show time costs of clean SFL and RISE as client number varies on CIFAR-10 and Fashion-MNIST.}}
    \label{fig:fidelity_avg_time}
\end{figure}

Fig.~\ref{fig:fidelity_main_task} reports main task accuracy, while Fig.~\ref{fig:fidelity_avg_time} shows training time costs under different numbers of clients on CIFAR-10 and Fashion-MNIST. The results indicate that watermarked models achieve comparable performance to clean models in both accuracy and efficiency. 
As shown in Tab.~\ref{tab:multi_model_multi_dataset_grand_table}, RISE consistently preserves fidelity across multiple network architectures, including ResNet-18 (RN18)~\cite{He2016ResNet18}, MobileNetV2 (MNV2)~\cite{Sandler2018MobileNetV2}, and DenseNet-121 (DN121)~\cite{Huang2017DenseNet121}, with accuracy differences within 2\% compared to clean SFL models. Meanwhile, it achieves watermark detectability exceeding 97\% across standard datasets in most cases. These results demonstrate that RISE maintains fidelity and generalizes well across model architectures.

\begin{table}[b]
\centering
\resizebox{\linewidth}{!}{
\begin{tabular}{llcccc}
\toprule
\textbf{Model} & \textbf{Dataset} & \textbf{Clean} & \textbf{$Acc_{\text{main}}$} & \textbf{$\theta_{\text{B}}$} & \textbf{$\theta_{\text{F}}$} \\
\midrule
\multirow{3}{*}{RN18} 
  & C10     & 81.84 & 81.78  & 99.32   & 100.00 \\
  & C100    & 50.05 & 49.36  & 99.62   & 100.00 \\
  & TINet   & 40.25 & 39.43  & 79.95   & 100.00 \\
\midrule
\multirow{3}{*}{MNV2} 
  & FM     & 91.21 & 90.32 & 100.00 & 100.00 \\
  & C10    & 85.14 & 84.99 & 99.97 & 100.00 \\
  & C100 & 53.91 & 53.36 & 69.90 & 100.00 \\
\midrule
\multirow{3}{*}{DN121} 
  & C10     & 84.24 & 83.07 & 99.90 & 100.00 \\
  & C100    & 52.26 & 51.49 & 98.10 & 100.00 \\
  & TINet & 47.51 & 44.04 & 97.15 & 100.00 \\
\bottomrule
\end{tabular}
}
\caption{RISE's performance of different models on various datasets.}
\label{tab:multi_model_multi_dataset_grand_table}
\end{table}

% \subsection{Robustness against Complex Dataset and Large Scalability}
% \noindent To further verify that the proposed RISE scheme is suitable for more complex situations, we conducted experiments on ImageNet-100 dataset and expanded the client population to 100 and 200 for CIFAR-10 dataset. 

\subsection{Robustness against Removal Attack}

% \noindent We evaluate the robustness of RISE against a series of watermark removal attacks, including fine-tuning, pruning and quantization. These attacks are conducted on a pair of RISE top and bottom models trained for 200 global epochs by 10 clients on CIFAR-10. 
\noindent We evaluate the robustness of RISE against common watermark removal attacks, including fine-tuning~\cite{Uchida2017watermarkDNN}, pruning~\cite{Lukas2022sokwm}, and quantization. These attacks are performed on RISE top and bottom models trained for 200 global epochs by 10 clients on CIFAR-10.

% pruning attack 实验折线图
\begin{figure}[t]
    \centering
    \begin{subfigure}[]{0.49\linewidth}
        \centering
        \includegraphics[width=\linewidth]{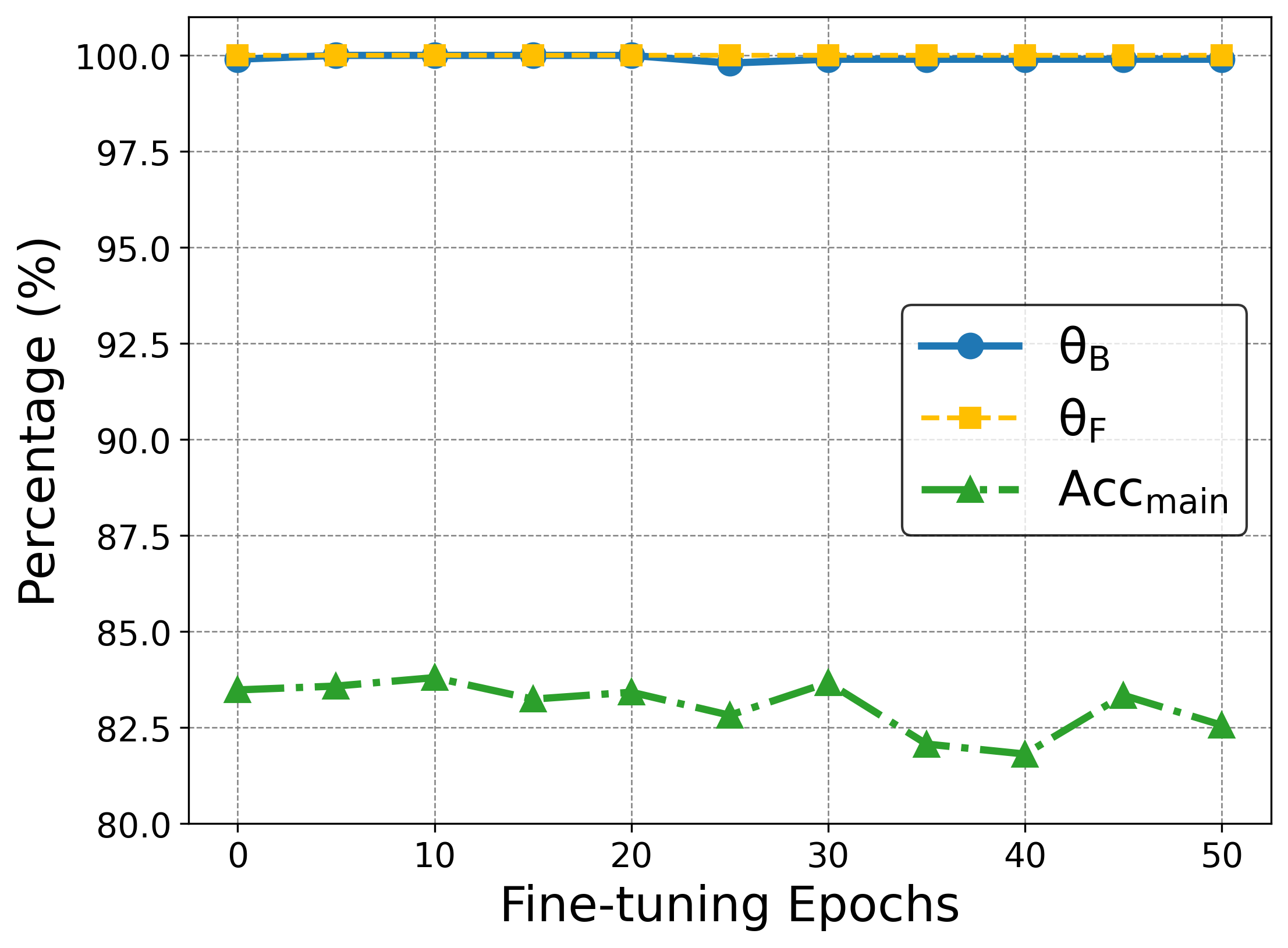}
        \caption{\textnormal{Fine-tuning Attack}}
        \label{fig:fine_tuning_attack}
    \end{subfigure}
    \hfill
    \begin{subfigure}[]{0.49\linewidth}
        \centering
        \includegraphics[width=\linewidth]{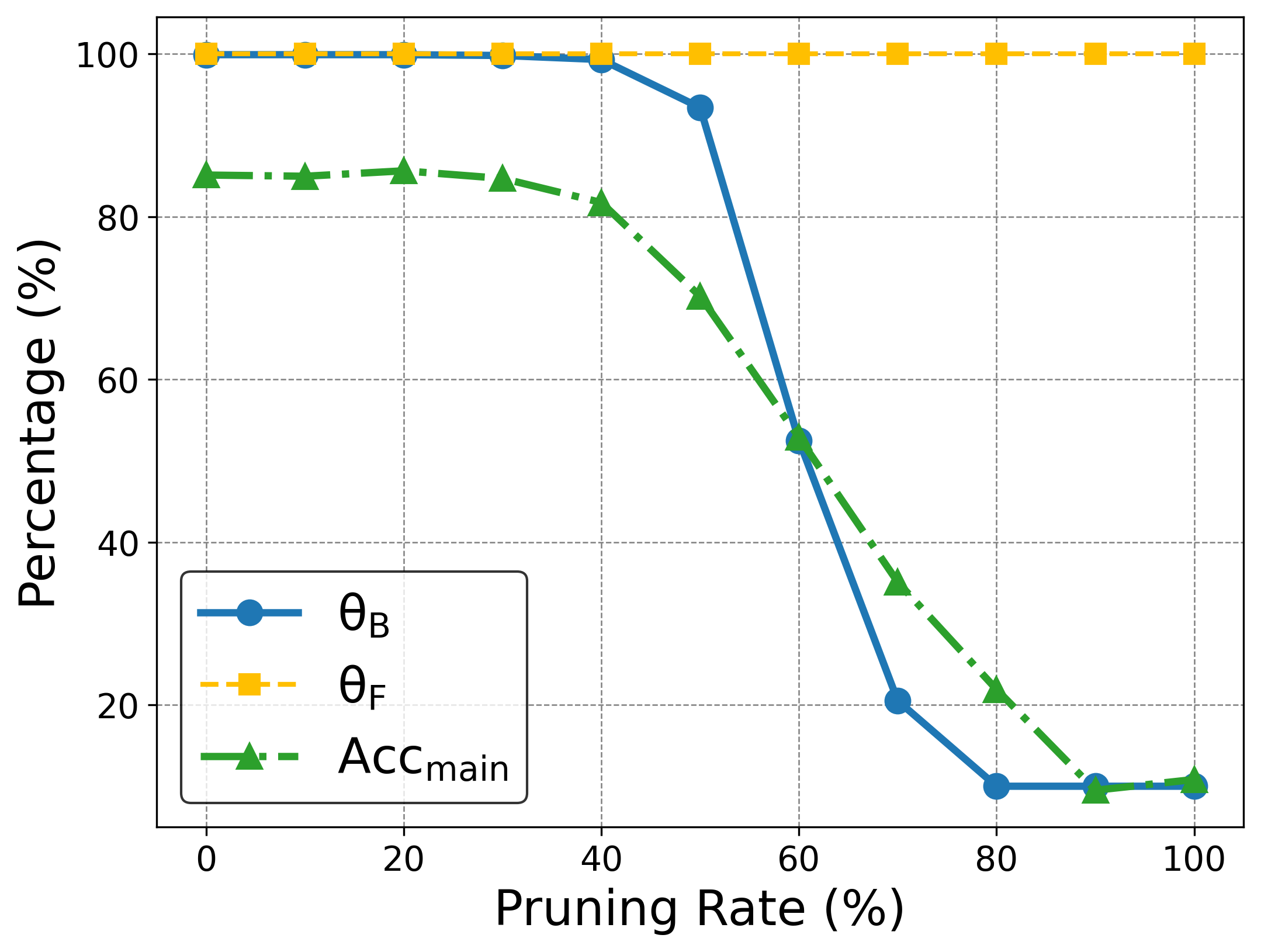}
        \caption{\textnormal{Pruning Attack}}
        \label{fig:pruning_attack}
    \end{subfigure}
    \caption{\textnormal{Figure presents main task and watermarking performance under fine-tuning and pruning attack on CIFAR-10 with ResNet18.}}
    \label{fig:ft_and_pruning}
\end{figure}

% \begin{figure}[t]
%     \begin{minipage}[t]{0.5\linewidth}
%         \centering
%         \includegraphics[width=\linewidth]{images/fine_tuning_diagram.png}
%         \caption{\textnormal{Robust Performance against Fine-tuning Attack}}
%         \label{fig:fine_tuning_attack}
%     \end{minipage}%
%     \hfill
%     \begin{minipage}[t]{0.49\linewidth}
%         \centering
%         \includegraphics[width=\linewidth]{images/pruning_rate_diagram_2.png}
%         \caption{\textnormal{Robust Performance against Pruning}}
%         \label{fig:pruning_attack}
%     \end{minipage}
% \end{figure}

\subsubsection{Robustness Against Fine-tuning Attack}

\noindent Fine-tuning attacks~\cite{Uchida2017watermarkDNN} attempt to erase watermarks by retraining the model with clean data, potentially overwriting neurons associated with watermarks. 

We simulate varying attacker capabilities by applying fine-tuning for 5 to 50 epochs. As shown in Fig.~\ref{fig:fine_tuning_attack}, both the server-side and client-side watermarking detection rates (\(\theta_F\) and \(\theta_B\)) remain stable, consistently above 99.80\%. The main task accuracy \(Acc_{\text{main}}\) fluctuates slightly between 81.81\% and 83.80\%. These results indicate that RISE is robust against fine-tuning attacks.

\subsubsection{Robustness Against Pruning Attack}

\noindent Pruning~\cite{Zhu2017Pruning} aims to remove redundant parameters from the trained model, which may erase embedded watermarks.

We assess $Acc_{\text{main}}$, $\theta_B$ and \(\theta_F\) under varying pruning rates. As shown in Fig.~\ref{fig:pruning_attack}, \(\theta_F\) remains unaffected at all pruning levels. The client-side watermarking detection rate \(\theta_B\) decreases noticeably as the pruning rate increases, but this drop is accompanied by a simultaneous decline in \(Acc_{\text{main}}\). When the pruning rate exceeds 60\%, although \(\theta_B\) drops sharply, the model becomes practically unusable for plagiarism, as \(Acc_{\text{main}}\) deteriorates to trivial levels. 

The results confirm that server-side watermarks are robust against pruning. Although client-side watermarks become less detectable as pruning intensifies, the model degrades to unusable, preventing plagiarism. Together, these findings suggest that RISE is robust against pruning attacks.

\subsubsection{Robustness Against Quantization Attack} 

\noindent Quantization~\cite{Hubara2017Quant, lin2018defensiveQuant} aims to remove watermarks by reducing numerical precision, e.g., converting floating-point parameters into integers, which may inadvertently eliminate watermark information.

% \begin{table}[t]
%     \centering
%     \begin{tabular}{lccc}
%     \toprule
%         \textbf{} & \textbf{float16} & \textbf{int32} & \textbf{int8} \\
%         \midrule
%         \(Acc_{\text{main}}\)  &\(80.88\% \pm 0.10\%\) & \(81.42\% \pm 0.43\%\) & \(63.86\% \pm 0.25\%\) \\
%         \(\theta_B\)    &\(99.30\% \pm 0.00\%\) & \(99.14\% \pm 0.05\%\) & \(92.36\% \pm 0.42\%\) \\
%         \(\theta_F\)    & \(100.00\% \pm 0.00\%\) & \(100.00\% \pm 0.00\%\) & \(100.00\% \pm 0.00\%\) \\
%         \bottomrule
%     \end{tabular}
%     \caption{\textnormal{Table shows the main task accuracy \(Acc_{\text{main}}\), the backdoor-based watermarking detection rate \(\theta_B\) and the feature-based watermarking detection rate \(\theta_F\) under 3 types of quantization attack, while the baseline achieves \(81.78\% \ Acc_{\text{main}}\), \(99.30\% \ \theta_B\) and \(100.00\% \ \theta_F\) in the CIFAR-10 classification task with \(K=10\) clients.}}
%     \label{tab:exp_quantization_without_baseline}
% \end{table}

\begin{table}[t]
    \centering
    \begin{tabular}{lcccc}
    \toprule
        \textbf{} & \textbf{baseline} & \textbf{FP16} & \textbf{INT32} & \textbf{INT8}  \\
        \midrule
        \(Acc_{\text{main}}\)  & \(81.78\%\) &\(80.88\%\) & \(81.42\%\) & \(63.86\%\) \\
        \(\theta_B\)  & \(99.32\%\)  &\(99.30\%\) & \(99.14\%\) & \(92.36\%\) \\
        \(\theta_F\)  & \(100.00\%\)  & \(100.00\%\) & \(100.00\%\) & \(100.00\%\) \\
        \bottomrule
    \end{tabular}
    \caption{\textnormal{RISE under 3 types of quantization attack on CIFAR-10.}}
    \label{tab:exp_quantization_without_baseline}
\end{table}

% As shown in Tab.~\ref{tab:exp_quantization_without_baseline}, quantization attacks to FP16 and INT32 slightly degrade the main task accuracy \(Acc_{\text{main}}\), while both \(\theta_B\) and \(\theta_F\) remain at 100\%. Under INT8 quantization, the server-side watermark remains stable, whereas \(\theta_B\) declines to 92.36\% and \(Acc_{\text{main}}\) drops to 63.86\%. These results suggest that RISE is robust against quantization attacks.

As shown in Tab.~\ref{tab:exp_quantization_without_baseline}, FP16 and INT32 quantization cause only minor main-task accuracy degradation, with both \(\theta_B\) and \(\theta_F\) remaining at 100\%. Under INT8 quantization, the server-side watermark remains stable, while \(\theta_B\) declines slightly to 92.36\% and \(Acc_{\text{main}}\) to 63.86\%. These results suggest that RISE is robust against quantization attacks.

\subsubsection{Robust against Neural Cleanse}
% \noindent Neural Cleanse (NC)~\cite{Wang2019NC} is a state-of-the-art backdoor detection technique that attempts to reverse-engineer trigger patterns via model's anomalous behaviors. In our experiments, we further leverage forged triggers to perform a trigger unlearning attack.
\noindent Neural Cleanse (NC)~\cite{Wang2019NC} is a state-of-the-art backdoor detection method that reverse-engineers candidate triggers by exploiting anomalous model behavior. 
We use forged triggers obtained via NC to launch a watermark unlearning attack against RISE.
% and the experiment result is shown as Tab.~\ref{tab:neural_cleanse}. 

% \begin{table}[h]
%     \centering
%     \begin{tabular}{cccc}
%         \toprule
%          \textbf{Detection Rate} & \textbf{False Positive} & \textbf{$\theta_{B}$} & \textbf{$\theta_F$} \\ 
%         \midrule
%         \(2.80\% \pm 6.94\%\) & 8.00\% & \(99.61\% \pm 0.63\%\) & 100.00\% \\
%         \bottomrule
%     \end{tabular}
%     \caption{\textnormal{Table describes the stealthiness and the robustness under Neural Cleanse backdoor detection}}
%     \label{tab:neural_cleanse}
% \end{table}

The average backdoor detection rate of NC is only \(2.80\% \pm 6.94\%\), showing that NC fails to effectively detect embedded watermarks in the SFL models. In addition, it produces \(8.00\%\) false positives by incorrectly flagging clean classes as backdoored. Notably, after the unlearning process, \(\theta_\text{B}\) remains high at \(99.61\% \pm 0.63\%\), and \(\theta_\text{F}\) maintains \(100.00\%\). 

These results indicate that RISE achieves strong stealthiness against NC, and is robust against the trigger unlearning attack based on mimic triggers.

\subsection{Comparison with FedIPR}

\noindent As the classic watermarking scheme FedIPR~\cite{Li2023FedIPR} in FL implemented both backdoor-based and feature-based watermarking, we adapt FedIPR to SFL following aforementioned ResNet-18 split scheme to demonstrate RISE's efficiency.

\begin{table}[b]
\centering
\resizebox{\linewidth}{!}{
\begin{tabular}{c cccccc}
\toprule
\multirow{2}{*}{\textbf{\(K\)}} & \multicolumn{2}{c}{$Acc_{\text{main}}$(\%)} & \multicolumn{2}{c}{$\theta_{\text{B}}$(\%)} & \multicolumn{2}{c}{$\theta_{\text{F}}$(\%)} \\
\cmidrule(lr){2-3} \cmidrule(lr){4-5} \cmidrule(lr){6-7}
& FedIPR & RISE & FedIPR & RISE & FedIPR & RISE \\
\midrule
10  & 79.30 & \textbf{81.78} & \textbf{100.00} & 99.32 & 100.00 & 100.00 \\
20  & 76.62 & \textbf{79.36} & 99.43 & \textbf{99.63} & 99.90 & \textbf{100.00} \\
30  & 74.41 & \textbf{77.85} & 89.29 & \textbf{98.97} & 99.90 & \textbf{100.00} \\
40  & 73.34 & \textbf{75.18} & 58.20 & \textbf{98.22} & 99.80 & \textbf{100.00} \\
50  & 72.67 & \textbf{74.55} & 67.06 & \textbf{93.87} & 95.40 & \textbf{100.00} \\
\bottomrule
\end{tabular}
}
\caption{Main Metrics of RISE and FedIPR on CIFAR-10}
\label{main_metrics_cmp_fedipr_cifar_10}
\end{table}

\begin{table}[htbp]
\centering
\resizebox{\linewidth}{!}{
\begin{tabular}{c cccccc}
\toprule
\multirow{2}{*}{\textbf{\(K\)}} & \multicolumn{2}{c}{$Acc_{\text{main}}$(\%)} & \multicolumn{2}{c}{$\theta_{\text{B}}$(\%)} & \multicolumn{2}{c}{$\theta_{\text{F}}$(\%)} \\
\cmidrule(lr){2-3} \cmidrule(lr){4-5} \cmidrule(lr){6-7}
& FedIPR & RISE & FedIPR & RISE & FedIPR & RISE \\
\midrule
10  & \textbf{91.88} & 91.51          & \textbf{100.00} & 99.32          & 100.00 & 100.00 \\
20  & 91.69          & \textbf{91.79} & 99.34           & \textbf{99.63} & 94.09  & \textbf{100.00} \\
30  & \textbf{91.35} & 91.22          & 75.75           & \textbf{98.97} & 68.75  & \textbf{100.00} \\
40  & \textbf{91.20} & 90.59          & 57.23           & \textbf{98.22} & 61.50  & \textbf{100.00} \\
50  & \textbf{90.91} & 90.24          & 31.25           & \textbf{93.87} & 58.50  & \textbf{100.00} \\
\bottomrule
\end{tabular}
}
\caption{Main Metrics Comparison between RISE and FedIPR on the Fashion-MNIST}
\label{main_metrics_cmp_fedipr_fashion}
\end{table}

We evaluate $Acc_{\text{main}}$, $\theta_{\text{B}}$ and $\theta_{\text{F}}$ in RISE and FedIPR. 
The experiments are conducted on CIFAR-10 and Fashion-MNIST under identical hyperparameters, with the number of clients varying. As shown in Tab.~\ref{main_metrics_cmp_fedipr_cifar_10} and Tab.~\ref{main_metrics_cmp_fedipr_fashion}, both approaches exhibit comparable performance in terms of main task accuracy. However, RISE achieves higher watermark detection rates whereas FedIPR demonstrates a noticeable degradation in watermark detectability as the number of clients increases. 

Furthermore, we conduct experiments on different model architectures with CIFAR-10 and CIFAR-100 to demonstrate RISE is more competent in the SFL scenario. As shown in Tab.~\ref{tab:main_metrics_cmp_fedipr_multi_scenario}, RISE performs slightly higher in main task accuracy \(Acc_{\text{main}}\). Meanwhile, RISE outperforms FedIPR in watermark detection rates \(\theta_{\text{B}}\) and \(\theta_{\text{F}}\) for MobileNetV2 and DenseNet-121.

% \begin{table}[t]
% \centering
% \resizebox{\linewidth}{!}{
% \begin{tabular}{c cccccc}
% \toprule
% \multirow{2}{*}{\textbf{Model}} & \multicolumn{2}{c}{$Acc_{\text{main}}$(\%)} & \multicolumn{2}{c}{$\theta_{\text{B}}$(\%)} & \multicolumn{2}{c}{$\theta_{\text{F}}$(\%)} \\
% \cmidrule(lr){2-3} \cmidrule(lr){4-5} \cmidrule(lr){6-7}
% & FedIPR & RISE & FedIPR & RISE & FedIPR & RISE \\
% \midrule
% RN18   & 79.30 & \textbf{81.78} & \textbf{100.00} & 99.32 & 100.00 & 100.00 \\
% MNV2   & 83.73 & \textbf{84.99} & 98.48 & \textbf{99.97} & 92.25 & \textbf{100.00} \\
% DN121  & 80.70 & \textbf{83.07} & 92.33 & \textbf{93.87} & 98.75 & \textbf{100.00} \\
% \bottomrule
% \end{tabular}
% }
% \caption{Main Metrics comparison of RISE and FedIPR in different models on CIFAR-10}
% \label{tab:main_metrics_cmp_fedipr_cifar_10_multi_models}
% \end{table}

\begin{table}[t]
\centering
\resizebox{\linewidth}{!}{
\begin{tabular}{cc cccccc}
\toprule
\multirow{2}{*}{\textbf{Model}} &
\multirow{2}{*}{\textbf{Dataset}} &
\multicolumn{2}{c}{$Acc_{\text{main}}$(\%)} &
\multicolumn{2}{c}{$\theta_{\text{B}}$(\%)} &
\multicolumn{2}{c}{$\theta_{\text{F}}$(\%)} \\
\cmidrule(lr){3-4} \cmidrule(lr){5-6} \cmidrule(lr){7-8}
& & FedIPR & RISE & FedIPR & RISE & FedIPR & RISE \\
\midrule
\multirow{2}{*}{RN18}
  & C10  & 79.30 & \textbf{81.78} & \textbf{100.00} & 99.32 & 100.00 & 100.00 \\
  & C100 & 48.93 & \textbf{49.36} & 98.20 & \textbf{99.62} & 82.33 & \textbf{99.95} \\
\midrule
\multirow{2}{*}{MNV2}
  & C10  & 83.73 & \textbf{84.99} & 98.48 & \textbf{99.97} & 92.25 & \textbf{100.00} \\
  & C100 & 50.15 & \textbf{53.36} & 50.60 & \textbf{69.90} & 90.25 & \textbf{100.00} \\
\midrule
\multirow{2}{*}{DN121}
  & C10  & 80.70 & \textbf{83.07} & 92.33 & \textbf{93.87} & 98.75 & \textbf{100.00} \\
  & C100 & 50.68 & \textbf{51.49} & 90.12 & \textbf{97.15} & 94.67 & \textbf{100.00} \\
\bottomrule
\end{tabular}
}
\caption{Comparison of RISE and FedIPR across different models and datasets.}
\label{tab:main_metrics_cmp_fedipr_multi_scenario}
\end{table}

Therefore, RISE is more suitable for the SFL architecture in terms of robustness and adaptability, especially when scaling to larger client numbers.

\subsection{Robustness against Free-rider Attack}

\noindent We evaluate whether RISE can resist free-riders---clients who exploit collaborative training without contributing genuine computation or data~\cite{Wang2022freerider}---by verifying the 
existence of client-side watermarks that indicate genuine participation.

Each client is required to disclose one of its private watermark patterns and declare the target misclassification. 
The client applies the trigger to a clean sample and submits both the original and modified inputs to the server. 
A successful misclassification into the declared class confirms the presence of a valid watermark and thus the client's participation. 
We consider two types of free-rider behaviors:
\begin{itemize}
    \item{\textbf{Type I Free-rider (Passive):}} Reuse the bottom model from the previous round without local updates.
    \item{\textbf{Type II Free-rider (Active):}} Use Neural Cleanse to forge a trigger and try to pass the verification.
\end{itemize}

% Neither type injects backdoored data during training, and thus lacks embedded watermarks to prove their participation.
% A client is marked as a free-rider if the declared trigger fails to induce the expected misclassification, indicating the absence of a valid client-side watermark.
Neither type injects backdoored data during training and therefore lacks embedded watermarks to prove participation. 
A client is identified as a free-rider if its declared trigger fails to induce the intended misclassification, indicating the absence of a legitimate client-side watermark.

\begin{figure}[htbp]
    \centering
    \begin{subfigure}[]{0.49\linewidth}
        \centering
        \includegraphics[width=\linewidth]{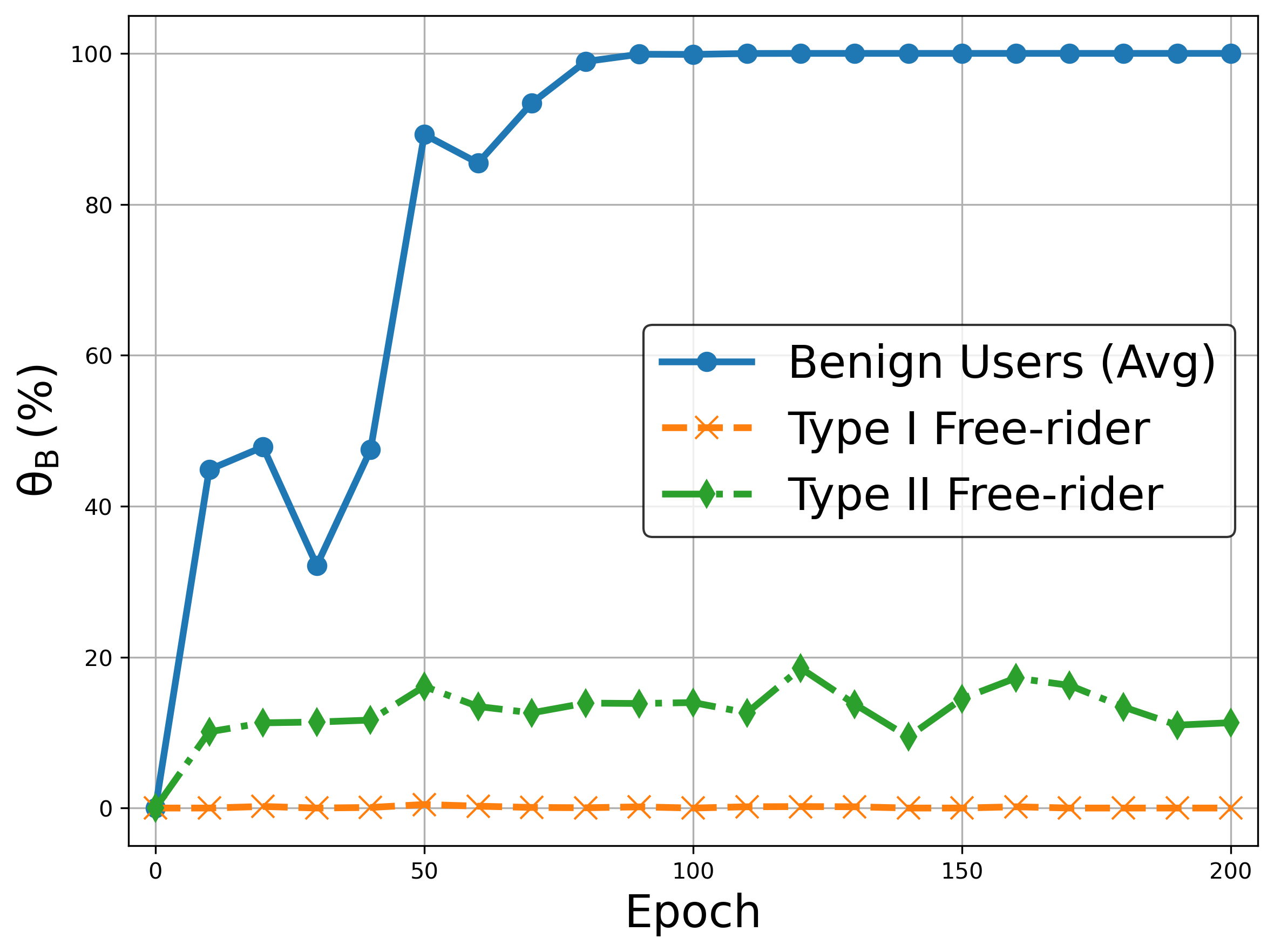}
        \caption{\textnormal{Freeriders attack}}
        \label{fig:freeriders_attack}
    \end{subfigure}
    \hfill
    \begin{subfigure}[]{0.49\linewidth}
        \centering
        \includegraphics[width=\linewidth]{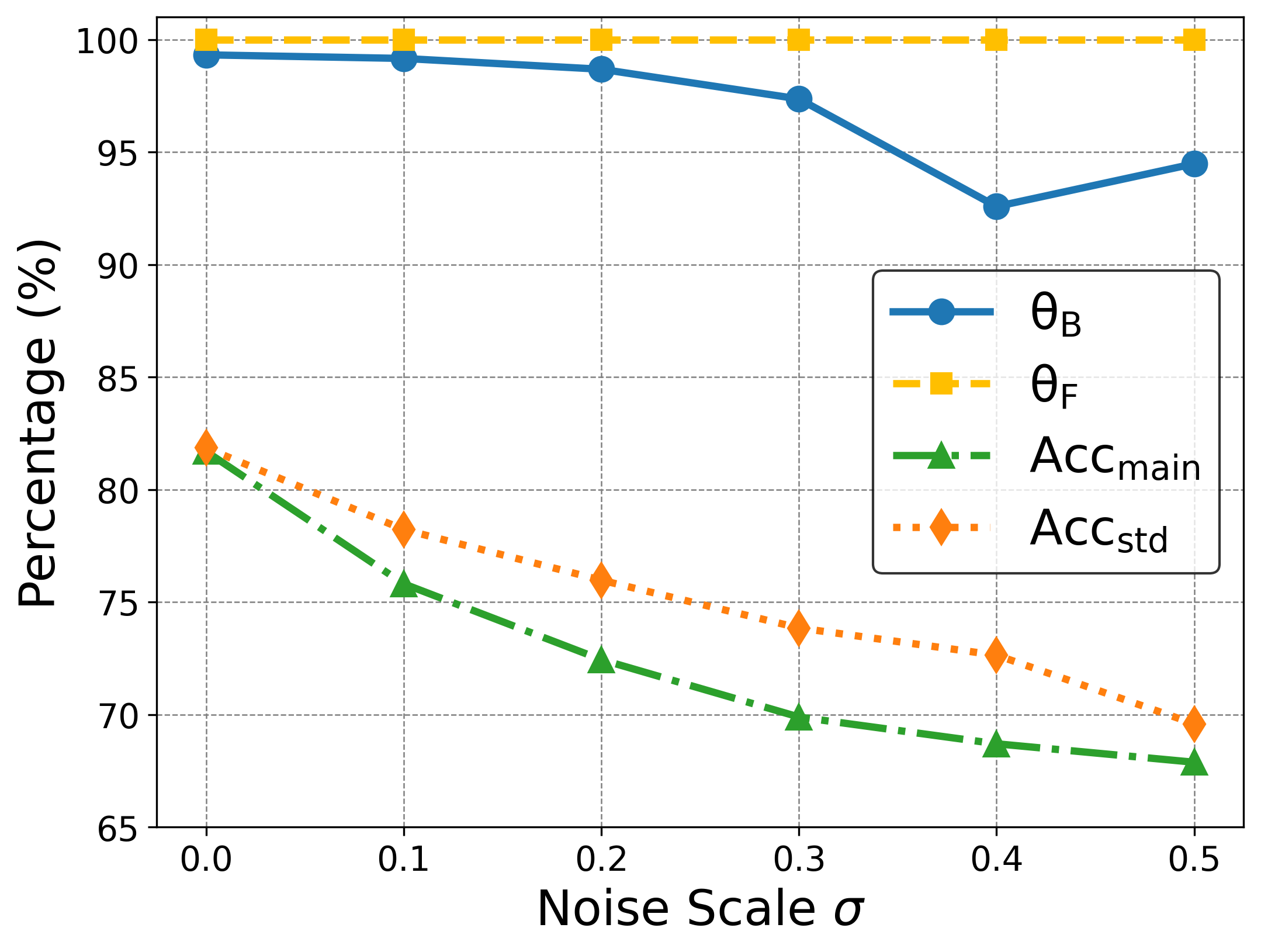}
        \caption{\textnormal{Differential Privacy}}
        \label{fig:dp_sigma}
    \end{subfigure}
    \caption{\textnormal{Freerider attack and differential privacy results on CIFAR-10 using ResNet18-based RISE.}}
    \label{fig:freeriders_and_dp}
\end{figure}

Experiments conducted with RISE on CIFAR-10 involve three benign clients, one Type~I free-rider, and one Type~II free-rider participating simultaneously. 
As shown in Fig.~\ref{fig:freeriders_attack}, the detection rate \(\theta_B\) for the Type~I free-rider remains near \(0\%\) throughout training, while that of the Type~II free-rider reaches only \(18.53\%\). 
These results demonstrate that neither type of free-rider can pass the watermark verification, indicating that RISE effectively resists free-rider attacks.
% As shown in Fig.~\ref{fig:freeriders_attack}, the \(\theta_B\) for the Type I free-rider is near 0\% throughout. For the Type II free-rider, \(\theta_B\) reaches up to \(18.53\%\). These results demonstrate that neither type of free-rider can pass the watermark verification. Therefore, RISE has the ability to defeat the free-rider attack.

% pruning attack 实验折线图

% % pruning attack 实验折线图
% \begin{figure}[t]
%     \begin{minipage}[t]{0.49\linewidth}
%         \centering
%         \includegraphics[width=\linewidth]{images/freeriders_avg_benign.png}
%         \caption{\textnormal{Robust Performance against Free-rider attack}}
%         \label{fig:freeriders_attack}
%     \end{minipage}%
%     \hfill
%     \begin{minipage}[t]{0.49\linewidth}
%         \centering
%         \includegraphics[width=\linewidth]{images/diagram_dp_sigma.png}
%         \caption{\textnormal{Robust Performance under Differential Privacy}}
%         \label{fig:dp_sigma}
%     \end{minipage}
% \end{figure}

\subsection{Robustness under Differential Privacy}

\noindent Adversaries can reconstruct client data from smashed data, threatening the privacy guarantees of SL~\cite{li2023gan, Xu2024FORA, Gao2023PCATFA}.

To mitigate this risk, RISE incorporates Differential Privacy (DP), enabling each client to specify an individual privacy budget. 
Since the bottom model processes raw inputs locally, directly adding noise at the input level may distort critical features and degrade performance.
Therefore, RISE injects noise into the smashed data instead, achieving a better trade-off between privacy protection and model utility.

We evaluate the impact of varying DP noise scales \(\sigma\) on RISE for CIFAR-10. Fig.~\ref{fig:dp_sigma} reports the main metrics of RISE, with \(Acc_{\text{std}}\) representing the main task accuracy of clean SFL under the same settings without watermarking.

The results show that \(\theta_F\) remains stable at \(100.00\%\), while \(\theta_B\) consistently exceeds 91.28\%. \(Acc_{\text{main}}\) drop compared to \(Acc_{\text{std}}\) stays within 5\%. These findings suggest that both sides of watermarking remain robust under DP, confirming the resilience of RISE in DP settings.

\subsection{Detailed Split Scheme}
\label{sec:detailed_split_scheme}
This section details the split schemes of RISE for different network architectures on CIFAR-10. 
To quantify the communication efficiency, we define the relative communication cost~\(\eta\) as the ratio between the cost of a given scheme and that of the minimal-cost scheme within the same architecture.

\subsubsection{Split Scheme for ResNet-18}

ResNet-18~\cite{He2016ResNet18} consists of an initial convolutional stem (\textit{conv stem}), 8 Residual Blocks (RB) and the classification head sequentially. The common splitting points are before and after each Residual Blocks~\cite{thapa2022splitfed}, as the classification head is trivial in terms of computation. However, in practical SFL scenarios, the server holds the classification head and at least one Residual Blocks. Thus, there are 8 splitting points in ResNet-18 for RISE. The ablation experiment is presented in Tab.~\ref{tab:split_ratio_results_resnet18}.

\begin{table*}[htbp]
\centering
\begin{tabular}{cccccccc}
\toprule
\textbf{Split} &
\textbf{Bottom Params} &
\textbf{Top Params} &
$\boldsymbol{Acc_{\text{main}}}$ (\%) &
$\boldsymbol{\theta_{\text{B}}}$ (\%) &
$\boldsymbol{\theta_{\text{F}}}$ (\%) &
$\boldsymbol{\phi}$ &
$\boldsymbol{\eta}$ \\
\midrule
1 (after stem)     & 9.5K   & 11.17M & 81.65 & 99.70  & 100.00 & 2.440 & 8.0 \\
2 (after 1st RB)   & 83.5K  & 11.09M & 81.78 & 99.32  & 100.00 & 2.320 & 8.0 \\
3 (after 2nd RB)   & 157.5K & 11.02M & 82.93 & 100.00 & 100.00 & 1.978 & 8.0 \\
4 (after 3rd RB)   & 387.6K & 10.80M & 83.12 & 100.00 & 100.00 & 1.601 & 4.0 \\
5 (after 4th RB)   & 683.1K & 10.50M & 83.66 & 99.90  & 100.00 & 1.405 & 4.0 \\
6 (after 5th RB)   & 1.602M & 9.580M & 84.12 & 99.60 & 100.00 & 1.139 & 2.0 \\
7 (after 6th RB)   & 2.783M & 8.399M & 83.46 & 99.90 & 100.00 & 0.596 & 2.0 \\
8 (after 7th RB)   & 6.456M & 4.726M & 84.62 & 99.80  & 100.00 & 0.350 & 1.0 \\
\bottomrule
\end{tabular}
\caption{Influence of different split points on task accuracy, watermark detection, time cost ratio $\phi$, and relative communication cost $\eta$ for ResNet-18 on CIFAR-10.}
\label{tab:split_ratio_results_resnet18}
\end{table*}

\subsubsection{Split Scheme for MobileNetV2}
\label{sec:split_scheme_for_mobilenet}

MobileNetV2~\cite{Sandler2018MobileNetV2} follows a stage-wise architecture composed of a lightweight convolutional stem (\textit{conv stem}), seven consecutive stages of inverted residual blocks (IRB), and a classification head that includes a $1\times1$ projection layer (\textit{conv last}), a global average pooling, and a linear classifier.

Each stage $s_i = (t_i, c_i, n_i, s_i)$ specifies the expansion ratio, output channels, number of blocks, and stride, respectively, as detailed in Tab.~\ref{tab:mobilenetv2_cfg}. 
For the split experiments, the network is partitioned at the boundaries between these seven stages, together with the positions before the \textit{conv stem} and after the \textit{conv last}, resulting in nine possible split configurations in total. 
The corresponding ablation results are summarized in Tab.~\ref{tab:split_ratio_results_mobilenet}.

\begin{table}[t]
\centering
\begin{tabular}{c|cccc}
\hline
\textbf{Stage} & $t$ & $c$ & $n$ & $s$ \\
\hline
1 & 1 & 16  & 1 & 1 \\
2 & 6 & 24  & 2 & 1 \\
3 & 6 & 32  & 3 & 2 \\
4 & 6 & 64  & 4 & 2 \\
5 & 6 & 96  & 3 & 1 \\
6 & 6 & 160 & 3 & 2 \\
7 & 6 & 320 & 1 & 1 \\
\hline
\end{tabular}
\caption{Configuration of the seven stages in MobileNetV2. Each stage $s_i=(t_i, c_i, n_i, s_i)$ denotes expansion factor $t$, output channels $c$, number of blocks $n$, and stride $s$.}
\label{tab:mobilenetv2_cfg}
\end{table}

\begin{table*}[htbp]
\centering
\begin{tabular}{cccccccc}
\toprule
\textbf{Split} &
\textbf{Bottom Params} &
\textbf{Top Params} &
$\boldsymbol{Acc_{\text{main}}}$ (\%) &
$\boldsymbol{\theta_{\text{B}}}$ (\%) &
$\boldsymbol{\theta_{\text{F}}}$ (\%) &
$\boldsymbol{\phi}$ &
$\boldsymbol{\eta}$ \\
\midrule
1 (after stem)     & 928    & 2.236M & 85.04 & 100.00 & 100.00 & 10.16  & 12.8 \\
2 (after 1st stage)   & 1824   & 2.235M & 84.99 & 100.00 & 100.00 & 8.108  & 6.4 \\
3 (after 2nd stage)   & 15.79K & 2.221M & 85.14 & 99.90  & 100.00 & 3.153  & 9.6 \\
4 (after 3rd stage)   & 55.49K & 2.181M & 85.38 & 100.00 & 100.00 & 2.178  & 3.2 \\
5 (after 4th stage)   & 239.4K & 1.997M & 86.61 & 100.00 & 100.00 & 1.440  & 1.6 \\
6 (after 5th stage)   & 542.5K & 1.694M & 87.37 & 100.00 & 100.00 & 0.8230 & 2.4 \\
7 (after 6th stage)   & 1.338M & 898.9K & 88.79 & 100.00 & 100.00 & 0.5340 & 1.0 \\
8 (after 7th stage)   & 1.812M & 425.0K & 89.24 & 100.00 & 100.00 & 0.3282 & 2.0 \\
\bottomrule
\end{tabular}
\caption{Influence of different split points on task accuracy, watermark detection, time cost ratio $\phi$, and relative communication cost $\eta$ for MobileNetV2 on CIFAR-10.}
\label{tab:split_ratio_results_mobilenet}
\end{table*}

\subsubsection{Split Scheme for DenseNet-121}
\label{sec:split_scheme_for_densenet}
DenseNet-121~\cite{Huang2017DenseNet121} consists of an initial convolutional stem (\textit{conv stem}), four densely connected blocks (DB1–DB4), and three transition layers in between, followed by a global average pooling and a linear classifier. 
Each dense block compRISEs multiple composite layers that concatenate feature maps from all preceding layers, promoting feature reuse and mitigating gradient vanishing. 
The transition layers perform $1\times1$ convolution and $2\times2$ average pooling to reduce spatial dimensions and control model complexity. 
For split experiments, the network is partitioned at the boundaries between the \textit{conv stem} and the four dense blocks (DB1–DB4), resulting in five split schemes in total.The ablation experiment is presented in Tab.~\ref{tab:split_ratio_results_densenet}.

\begin{table*}[!t]
\centering
\begin{tabular}{cccccccc}
\toprule
\textbf{Split} &
\textbf{Bottom Params} &
\textbf{Top Params} &
$\boldsymbol{Acc_{\text{main}}}$ (\%) &
$\boldsymbol{\theta_{\text{B}}}$ (\%) &
$\boldsymbol{\theta_{\text{F}}}$ (\%) &
$\boldsymbol{\phi}$ &
$\boldsymbol{\eta}$ \\
\midrule
1 (after stem)     & 928    & 2.236M & 81.99 & 100.00 & 100.00 & 10.16  & 12.8 \\
2 (after DB1)   & 1824   & 2.235M & 83.07 & 100.00 & 100.00 & 8.108  & 6.4 \\
3 (after DB2)   & 15.79K & 2.221M & 84.33 & 99.90  & 100.00 & 3.153  & 9.6 \\
4 (after DB3)   & 55.49K & 2.181M & 86.11 & 100.00 & 100.00 & 2.178  & 3.2 \\
5 (after DB4)   & 239.4K & 1.997M & 86.20 & 100.00 & 100.00 & 1.440  & 1.6 \\
\bottomrule
\end{tabular}
\caption{Influence of different split points on task accuracy, watermark detection, time cost ratio $\phi$, and relative communication cost $\eta$ for DenseNet-121 on CIFAR-10.}
\label{tab:split_ratio_results_densenet}

\end{table*}
\section{Conclusion}
\noindent In this paper, we proposed RISE, a robust intellectual property protection scheme designed for SFL. Unlike traditional client-side or server-side watermarking schemes, RISE introduces a joint client-server watermark embedding mechanism, allowing both the server and clients to independently verify model ownership. The server embeds feature-based watermark into top-layer parameters via a watermarking regularization term, while clients inject backdoor-based watermarks into their local datasets using predefined triggers. Our experimental results validate the effectiveness of RISE, demonstrating its high fidelity, strong watermark detection significance, and robustness against various removal attacks.

% In future work, we aim to extend RISE beyond ownership verification by incorporating traceability mechanisms that enable the identification of unauthorized model usage and distribution. This enhancement will further strengthen intellectual property protection in SFL and mitigate risks associated with model misuse.

In future work, we plan to extend RISE beyond ownership verification by incorporating traceability mechanisms to identify unauthorized model usage and distribution. This enhancement will further strengthen intellectual property protection in SFL and reduce the risk of model misuse.

% Future work will extend RISE from ownership verification to traceability, enabling the detection of unauthorized model use and improving IP protection in SFL.
\newpage
{
    \small
    \bibliographystyle{ieeenat_fullname}
    \bibliography{main}
}


\end{document}